\documentclass[fleqn,usenatbib]{mnras}

\usepackage{enumitem}

\usepackage[T1]{fontenc}

\DeclareRobustCommand{\VAN}[3]{#2}
\let\VANthebibliography\thebibliography
\def\thebibliography{\DeclareRobustCommand{\VAN}[3]{##3}\VANthebibliography}

\usepackage{graphicx}	
\usepackage{amsmath}	
\usepackage{amssymb}	
\usepackage{color}
\usepackage{ulem}
\usepackage{comment}
\usepackage{multirow}
\usepackage{newtxtext,newtxmath}
\usepackage{caption}

\newcommand\cyanuline{\bgroup\markoverwith{\textcolor{cyan}{\rule[0.5ex]{2pt}{1.6pt}}}\ULon}

\newcommand{\cp}{\mathrm{c_\mathnormal{\phi}}}
\renewcommand{\sp}{\mathrm{s_\mathnormal{\phi}}}
\newcommand{\ct}{\mathrm{c_\mathnormal{\theta}}}
\newcommand{\st}{\mathrm{s_\mathnormal{\theta}}}
\newcommand{\cpp}{\mathrm{c_\mathnormal{\phi'}}}
\newcommand{\spp}{\mathrm{s_\mathnormal{\phi'}}}
\newcommand{\ctp}{\mathrm{c_\mathnormal{\theta'}}}
\newcommand{\stp}{\mathrm{s_\mathnormal{\theta'}}}

\title[]{Jet-environment interplay in magnetized binary neutron star mergers}

\author[A. Pavan et al.]{
Andrea Pavan,$^{1,2,3}$\thanks{E-mail: andrea.pavan.20@phd.unipd.it (AP)}
Riccardo Ciolfi,$^{2,3}$\thanks{E-mail: riccardo.ciolfi@inaf.it (RC)}
Jay V. Kalinani$^{2,3}$
and Andrea Mignone$^{4}$
\\
$^{1}$Dipartimento di Fisica e Astronomia, Universit\`a di Padova, Via Francesco Marzolo 8, I-35131 Padova, Italy\\
$^{2}$INAF, Osservatorio Astronomico di Padova, Vicolo dell'Osservatorio 5, I-35122 Padova, Italy\\
$^{3}$INFN, Sezione di Padova, Via Francesco Marzolo 8, I-35131 Padova, Italy\\
$^{4}$Dipartimento di Fisica, Universit\`a di Torino, Via Pietro Giuria 1, I-10125 Torino, Italy
}

\date{Accepted XXX. Received YYY; in original form ZZZ}

\pubyear{2022}

\begin{document}
\label{firstpage}
\pagerange{\pageref{firstpage}--\pageref{lastpage}}
\maketitle

\begin{abstract} 
GRB\,170817A, the first short gamma-ray burst (sGRB) to be detected in coincidence with a gravitational wave signal, demonstrated that merging binary neutron star (BNS) systems can power collimated ultra-relativistic jets and, in turn, produce sGRBs. Moreover, it revealed that sGRB jets possess an intrinsic angular structure that is imprinted in the observable prompt and afterglow emission. 
Advanced numerical simulations represent the leading approach to investigate the physical processes underlying the evolution of sGRB jets breaking out of post-merger environments, and thus connect the final angular structure and energetics with specific jet launching conditions. 
In a previous paper, we carried out the first three-dimensional (3D) special-relativistic hydrodynamic simulations of incipient (top-hat) sGRB jets propagating across the realistic environment resulting from a general-relativistic (GR) hydrodynamic BNS merger simulation. 
While the above work marked an important step toward a consistent end-to-end description of sGRB jets from BNS mergers, those simulations did not account for the presence of magnetic fields, which are expected to play a key role. 
Here, we overcome this limitation, reporting the first 3D special-relativistic magnetohydrodynamic (MHD) simulation of a magnetized (structured and rotating) sGRB jet piercing through a realistic magnetized post-merger environment, wherein the initial conditions of the latter are directly imported from the outcome of a previous GRMHD BNS merger simulation. 
\end{abstract}

\begin{keywords} 
MHD -- gamma-ray bursts -- stars: jets -- neutron star mergers -- relativistic processes -- methods: numerical.
\end{keywords}

\section{Introduction}
\label{intro}

The association between binary neutron star (BNS) mergers and short gamma-ray bursts (sGRBs), put forward over three decades ago \citep[e.g.,][]{Paczynski1986,Eichler1989}, was recently confirmed via the first joint detection of gravitational waves (GWs) from the coalescence of two NSs (event named GW170817) and a burst of gamma-rays (GRB\,170817A), occurred in August 2017 (\citealt{LVC-BNS,LVC-Hubble,LVC-MMA,LVC-GRB}; see, e.g., \citealt{Margutti2021} for a recent review). 
A direct link between the two phenomena was definitely established hundreds of days later through accurate analysis of the sGRB afterglow emission, which revealed the presence of a structured, narrowly collimated ultra-relativistic jet fed by the merger, observed about 20-25 degrees away from its main propagation axis \citep[][and refs.~therein]{Mooley2018b,Ghirlanda2019,Mooley2022}. This unusual sGRB detection provided compelling evidence for the existence of an intrinsic angular structure of relativistic outflows powered by BNS mergers, further stimulating an in-depth investigation of the involved physical mechanisms \citep[][and refs.~therein]{Salafia2022}. 

Theoretical research focussing on the evolution of sGRB jets, from the initial launch to the final emerging structure and energetics, is continuously improving our understanding of the GRB\,170817A prompt and afterglow signals. In this context, high-resolution numerical simulations are playing a pivotal role. 
In particular, numerous studies based on jet simulations within a special- or general-relativistic framework, in hydro- or magnetohydro-dynamics, and in two or three dimensions, already investigated the role of different system features in determining the ultimate angular structure and observational signatures \citep[e.g.,][]{Lazzati2018,Xie2018,Kathirgamaraju2019,Geng2019,Nathanail2020,Nathanail2021,Murguia2021,Urrutia2021,Urrutia2023,Gottlieb2020,Gottlieb2021,Gottlieb2022a}.

The above effort in reproducing the jet evolution in a more and more realistic fashion is however affected by the common limitation that the post-merger matter distribution through which the incipient jet propagates is set via simple, hand-made prescriptions, lacking a direct connection to the self-consistent environment of any specific BNS merger.
As a notable exception, \cite{Nativi2021} \citep[see also][]{Nativi2022,Lamb2022} used an initial setup reproducing the outcome of a three-dimensional (3D) Newtonian simulation of an unmagnetized neutrino-driven wind emerging from a massive neutron star merger remnant.

As a first step toward a more consistent description overcoming the above limitation, in \cite{Pavan2021} we presented the first study of sGRB jet propagation inside `realistic' surrounding environments, i.e. directly imported from the outcome of merging BNS systems. 
Specifically, we carried out 3D special-relativistic hydrodynamic simulations of uniform incipient jets, manually injected into the environment resulting from a fully general-relativistic (hydrodynamic) BNS merger simulation, also performed by our group. 
Our study probed the effects of deviations from axisymmetry and homologous expansion as well as anisotropies, which characterize more realistic matter distributions. 
We found that, under the same jet injection conditions, hydrodynamic jets propagating across such a post-merger environment emerge with a significantly different breakout time, energy partitioning, and final angular structure with respect to the equivalent simulation with a typical simplified environment. 

Our work was followed by other jet propagation studies based on data that were directly imported from BNS merger simulations.  
In particular, \cite{Lazzati2021} also performed a relativistic hydrodynamic simulation of a jet injected into a realistic post-merger environment, showing that inhomogeneities in the latter, which cause oscillations of the jet's centroid around the main propagation axis, may allow the jet to propagate easier and faster compared to an identical jet in a smooth medium \citep[in agreement with results by][]{Pavan2021}. 

For a further quality step in our description of sGRB jets, magnetic fields represent the next crucial ingredient to be included.
Different studies have indeed shown that magnetic fields play a key role in the jet breakout, collimation, and further propagation \citep[e.g.,][]{Nathanail2020,Nathanail2021,Gottlieb2022a,Gottlieb2022b}. However, any inference on this has been so far obtained by taking into account only the magnetic field of the jet, without considering the (potentially high) magnetization of the environment in which jet propagation takes place.\footnote{We note that, most recently, \citet{Garcia-Garcia2023} presented 2.5\,D simulations of non-magnetized jets propagating through an idealized environment endowed with a simple poloidal magnetic field.
In that work, the authors discuss the impact of such an ambient magnetic field on the jet evolution, considering different field strengths and, correspondingly, different environment magnetizations.}

In the present work, we report the first 3D special-relativistic magneto-hydrodynamic (RMHD) simulation of an incipient magnetized sGRB jet propagating across the realistic environment resulting from a magnetized BNS merger.
We discuss the specific treatment for importing the magnetic field components from the reference merger simulation and provide an advanced and physically motivated prescription for injecting magnetized and rotating incipient jets, inspired by \cite{Geng2019} \citep[also see][]{Marti2015}. 
In addition, we develop a method to extend the explorable range of jet launching times using in full the information provided by the original BNS merger. 
Then, we present the results of our fiducial simulation extending up to $\mathrm{\simeq\!2\,s}$ after jet launching and analyze various aspects of the jet-environment interaction and the final jet structure.

The paper is organized as follows. 
In Section\,\ref{setup}, we describe our simulation setup, including numerical methods, data import procedure, grid structure, and prescriptions adopted for the jet injection. 
In Section\,\ref{extr}, we discuss the method that allows for a wider range of jet launching times.
In Section\,\ref{fiducial}, we present our fiducial model and the corresponding results. 
Finally, in Section\,\ref{summary}, we give a summary of the work and concluding remarks.

\section{Simulation setup}
\label{setup}

We perform our simulations using the special-relativistic module of the \textsc{PLUTO} code \citep{Mignone2007-PLUTO1,Mignone2012-PLUTO2}, which allows us to solve the equations of RMHD in multiple spatial dimensions and different geometries. 
In particular, we carry out our simulations using piecewise parabolic reconstruction, the Harten-Lax-van Leer (HLL) Riemann solver and third-order Runge Kutta time stepping in 3D spherical coordinates $(r,\theta,\phi)$.
Additionally, to enforce the free-divergence constraint of the magnetic field, we employ the Hyperbolic Divergence Cleaning
technique of \cite{Dedner2002} \citep[also see][]{MT2010,MTB2010}, which allows us to retain a cell-centered representation of the primary fluid variables, including the magnetic field, in our simulations. 
Details on divergence cleaning tests and the chosen setup are given in Appendix\,\ref{cleaning}.

As in \citet[henceforth P21]{Pavan2021}, we perform our simulations using initial data that are directly imported from the outcome of a general-relativistic (GR) BNS merger simulation.
Specifically, we remap the results of such simulation into \textsc{pluto}, setting up the computational domain with an inner radial excision of 380\,km radius to safely neglect GR effects (not accounted for by \textsc{pluto}) in our calculations. To reproduce the matter and magnetic field fluxes on the spherical surface $r_{\mathrm{exc}}\!=\!380$\,km, we set up appropriate radial boundary conditions that are based on the results of the reference BNS merger simulation. In Section\,\ref{import}, we describe in detail our import procedure and the resulting initial setup of our \textsc{pluto} simulations. The boundary condition settings at $r\!=\!r_{\mathrm{exc}}$ are instead specified in the following Sections \ref{collapse} and \ref{jet} (also see Section\,\ref{extr}).

The reference BNS merger simulation was performed in 3D-GRMHD, and corresponds to the `B5e15' model of \cite{Ciolfi2020a}. 
This was based on a BNS system that has the same chirp mass as estimated for GW170817 \citep{LVC-170817properties} and a mass-ratio $\mathrm{\simeq\!0.9}$. Moreover, the two initial NSs were endowed with an internal, purely poloidal magnetic field of maximum strength $\mathrm{5\!\times\!10^{15}\,G}$ and the equation of state (EOS) was a piecewise-polytropic approximation of the APR4 EOS \citep{Akmal:1998:1804} as implemented in \cite{Endrizzi2016}. The high magnetic field strength adopted for the two NSs was set to achieve the large amount of magnetic energy expected after the merger ($\mathrm{\simeq\!10^{51}\,erg}$), despite the lack of sufficient resolution to fully capture the main small-scale magnetic field amplification mechanisms during and after merger \cite[see][and references therein]{Ciolfi2020a}. 
The reference BNS merger simulation was performed on a hierarchical 3D-Cartesian computational grid, with seven refinement levels (finest spacing of $\mathrm{\approx\!250\,m}$) and all axes extended to $\mathrm{\approx\!3400\,km}$. Reflection symmetry was imposed on the equatorial plane (i.e., only the $z\!\geq\!0$ region was evolved).
For numerical stability, a constant density floor of $\mathrm{\rho^{*}\!=\!6.3\!\times\!10^4\,g/cm^3}$ was imposed, corresponding to a total mass of $\mathrm{\simeq\!3.5\!\times\!10^{-3}\,\mathnormal{M}_{\odot}}$ within a sphere of 3000\,km radius.
Further details about numerical codes and methods can be found in \cite{Ciolfi2017,Ciolfi2019} and \cite{Ciolfi2020a}.

The above simulation led to the formation of a magnetized and differentially rotating metastable NS remnant with mass $M_0\!\simeq\!2.596\,M_{\odot}$, surrounded by an expanding dense cloud of material (only in minor part unbound). 
The evolution was followed up to $\mathrm{\simeq\!255\,ms}$ after merger, and no collapse into a black hole (BH) occurred within such a timespan.
A mildly-relativistic ($\mathrm{\Gamma\!\lesssim\!1.05}$) collimated outflow fed by the central NS was observed during the simulation (see Figure\,\ref{fig1} below). However, considerations on energetics and outflow velocity ($\lesssim\!0.2\,c$) excluded the possibility that a typical sGRB jet could be produced \citep[see][]{Ciolfi2020a}. 

In our present study, we adopt the scenario in which a sGRB jet is launched by the accreting BH formed after the eventual collapse of a hypermassive NS remnant.\footnote{This the leading scenario to produce sGRB jets in BNS mergers and finds support in current BNS merger simulations \citep[e.g.,][]{Ruiz2016,Ciolfi2020a,Sun2022}.}
As in P21, we treat the time of collapse $t_\mathrm{c}$ as a parameter to be explored.  
We import data from the BNS merger simulation at a certain time after merger (see the next Section). 
Then, at a given physical time after data import, we assume that the NS remnant collapses into a BH, forming an accreting system that launches a sGRB jet after a few tens of ms (in this work, 30\,ms; see below). 
Within the latter time window, hereafter referred to as the `collapse phase', we mimic the effects of the collapse itself on the evolution of the surrounding environment before the emergence of an incipient jet (see Section\,\ref{collapse}). Finally, we proceed with the jet injection using the prescription detailed in Section\,\ref{jet} (also see Appendix\,\ref{jet_calc}), and follow the jet evolution up to $\mathrm{\sim\!2\,s}$ after injection.\footnote{While here the jet is injected by hand, future BNS merger simulations will provide us with self-consistent incipient jets that can be imported directly into \textsc{pluto} to study their following propagation on large scales.} 

To carry out our simulations in \textsc{pluto}, we set up the computational grid with logarithmic and uniform spacing along the radial and angular directions, respectively, and resolution of $N_r\!\times\!N_{\theta}\!\times\!N_{\phi}\!=\!768\!\times\!256\!\times\!512$ cells along $r,\theta,$ and $\phi$, respectively (smallest grid spacing, at $r_\mathrm{exc}$, of $\mathrm{\Delta\mathnormal{r}\!\simeq\!4.4\,km}$, $\mathrm{\mathnormal{r}\Delta\mathnormal{\theta}\!\simeq\!4.4\,km}$, and $\mathrm{\mathnormal{r}\Delta\mathnormal{\phi}\!\simeq\!4.7\,km}$).\footnote{We note that full numerical convergence would require higher resolution (see also discussion in P21). Demonstrating convergence and obtaining quantitatively accurate results, for a proper comparison with observational data (e.g., GRB\,170817A), will be the goal of our future work.}
We also restrict $\theta$ within the interval $\mathrm{[0.1,\pi-0.1]}$ to remove the polar axis singularity. Moreover, we set zero-gradient (or `outflow') and periodic boundary conditions for the $\theta$ and $\phi$ coordinates, respectively, and user-defined radial boundary conditions at $r=r_\mathrm{exc}$ (see below). At the outer radial boundary (i.e., at $r\!=\!r_\mathrm{max}\!=\!2.5\times10^6$\,km), eventually, we set outflow boundary conditions, allowing matter to flow naturally out of the computational domain.

In the following, we describe our import procedure (including the initial conditions setting), the modelling of the collapse phase, and the prescription employed for the jet injection.

\subsection{Data import}
\label{import}
\begin{figure}
\centering
	\includegraphics[width=0.78\columnwidth,keepaspectratio]{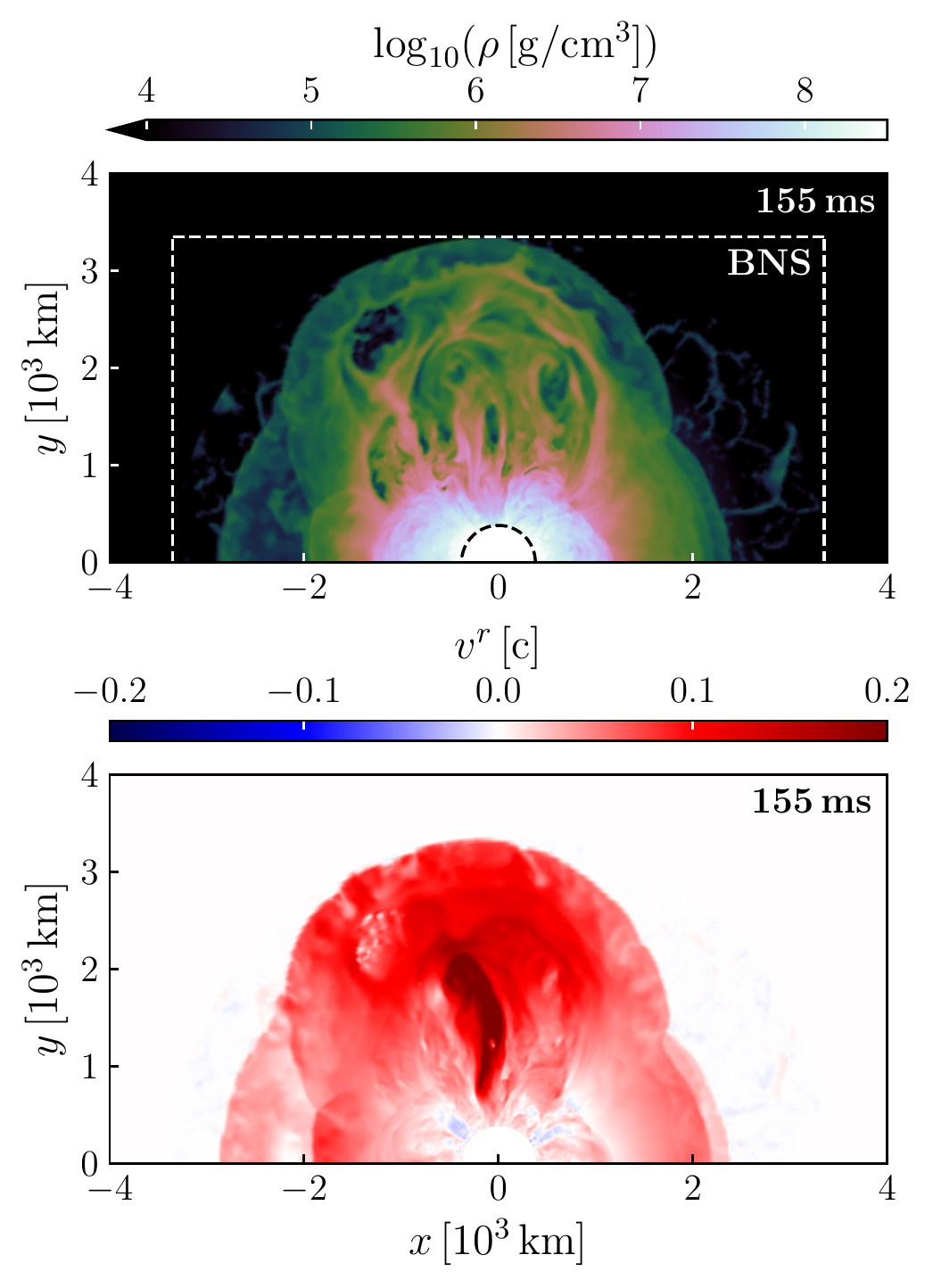}
    \caption{Meridional view of rest-mass density (top) and radial velocity (bottom) for data imported at 155\,ms after merger (see text). 
    In the upper plot, the white region at the center, bordered by a dashed-black line, corresponds to the excised region not evolved in \textsc{pluto}. Within the same plot, the dashed-white contour corresponds to the outer grid boundary of the the reference BNS merger simulation.}
    \label{fig1}
\end{figure}

The reference BNS merger simulation provides 3D outputs of rest-mass density, pressure, 3-velocity, and magnetic field. These are saved every $\mathrm{\simeq\!10\,ms}$ during the evolution, enforcing the equatorial symmetry with respect to the $xy$ plane. To import the relative data into \textsc{pluto}, we follow the procedure already presented in P21, with the addition of specific treatment for the magnetic fields. In particular:

(i) we remap data from the original domain to a uniform Cartesian one, with same resolution at the radial distance of 380\,km (remapping onto the uniform grid is performed using the publicly available \textsc{PostCacuts} package; \hyperlink{https://github.com/wokast/PyCactus}{https://github.com/wokast/PyCactus});

(ii) we compute values on both $\mathrm{\mathnormal{z}\!>\!0}$ and $\mathrm{\mathnormal{z}\!<\!0}$ by exploiting the equatorial symmetry of the data;

(iii) we tilt the imported system by $90^{\circ}$ in order to arrange the orbital axis orthogonally to the coordinate $z$-axis;\footnote{Tilting the system allows us to avoid dealing with the polar axis singularity when injecting the jet along the orbital axis of the system in 3D spherical coordinates (see Section\,\ref{jet}).}

(iv) we interpolate data from the uniform Cartesian domain to the spherical one employed in \textsc{pluto}. 

In this work, data import is performed at $\mathrm{\mathnormal{t}_0\!\simeq\!155\,ms}$ after merger, when the post-merger ejecta have almost reached the outer boundary of the Cartesian domain (see Figure\,\ref{fig1}). In such a way, we avoid losing any information related to the expanding material.

To relate thermodynamic quantities, we employ the Taub EOS \citep[see][and refs.~therein]{Mignone2007}, which corresponds to an ideal gas EOS with adiabatic index $\mathrm{\Gamma_{\mathrm{ad}}\!=\!4/3}$ in the highly relativistic limit, and $\mathrm{\Gamma_{\mathrm{ad}}\!=\!5/3}$ in the non-relativistic one, with smooth and continuous behaviour at intermediate regimes. 
Such an EOS allows us to properly describe our initial system, which is essentially non-relativistic, and to later account for an ultra-relativistic component (i.e.~the sGRB jet). 
However, we note that it may be defective in reproducing the radiation-mediated shocks formed during the jet propagation in the surrounding post-merger environment \citep{LevisonNakar2020}, for which a simpler ideal gas EOS with $\mathrm{\Gamma_{\mathrm{ad}}\!=\!4/3}$ would represent a better choice \citep[see, e.g.,][]{Gottlieb2022a}. For completeness, we performed our simulations also with the latter EOS, and in Appendix\,\ref{IDEALvsTAUB} we provide an in-depth comparison with the jet evolution given by the Taub EOS.

We further note, as already discussed in P21, that the Taub EOS does not exactly match the one used in the reference BNS merger simulation (within the density range of interest).
In order to quantify the impact of such a mismatch on the dynamical evolution of the system, in P21 we performed twice our fiducial simulation where either (i) the pressure was directly imported and the specific internal energy was derived from the Taub EOS or (ii) the opposite.
Here, we carried out the analogous test for our magnetized fiducial model, and the results are reported in Appendix\,\ref{Mismatch}.

When interpolating onto the \textsc{pluto}'s spherical grid, we substitute the constant density floor $\mathrm{\rho^{*}\!=\!6.3\!\times\!10^4\,g/cm^3}$ adopted in the reference BNS merger simulation with a new one that is more suitable for our investigation. In particular, we keep the original floor up to the radial distance of $\mathrm{2000\,km}$, to maintain numerical stability in the highest density region of the system, and from that radius on we replace it with a static, unmagnetized `atmosphere' with density and pressure decaying as $r^{-\alpha}$ up to $r_\mathrm{max}$. 
Moreover, we change the prescription adopted in P21 from $\mathrm{\alpha\!=\!5}$ to $\mathrm{\alpha\!=\!6.5}$, lowering by two orders of magnitude the mass of the atmosphere above the jet breakout radius ($r_\mathrm{b}\!\simeq\!5\!\times\!10^4$\,km, see Section\,\ref{fiducial}).
As a result, any potential atmospheric effect on the jet evolution that could be present with the P21 prescription is now strongly suppressed.
\begin{figure}
	\includegraphics[width=\columnwidth,keepaspectratio]{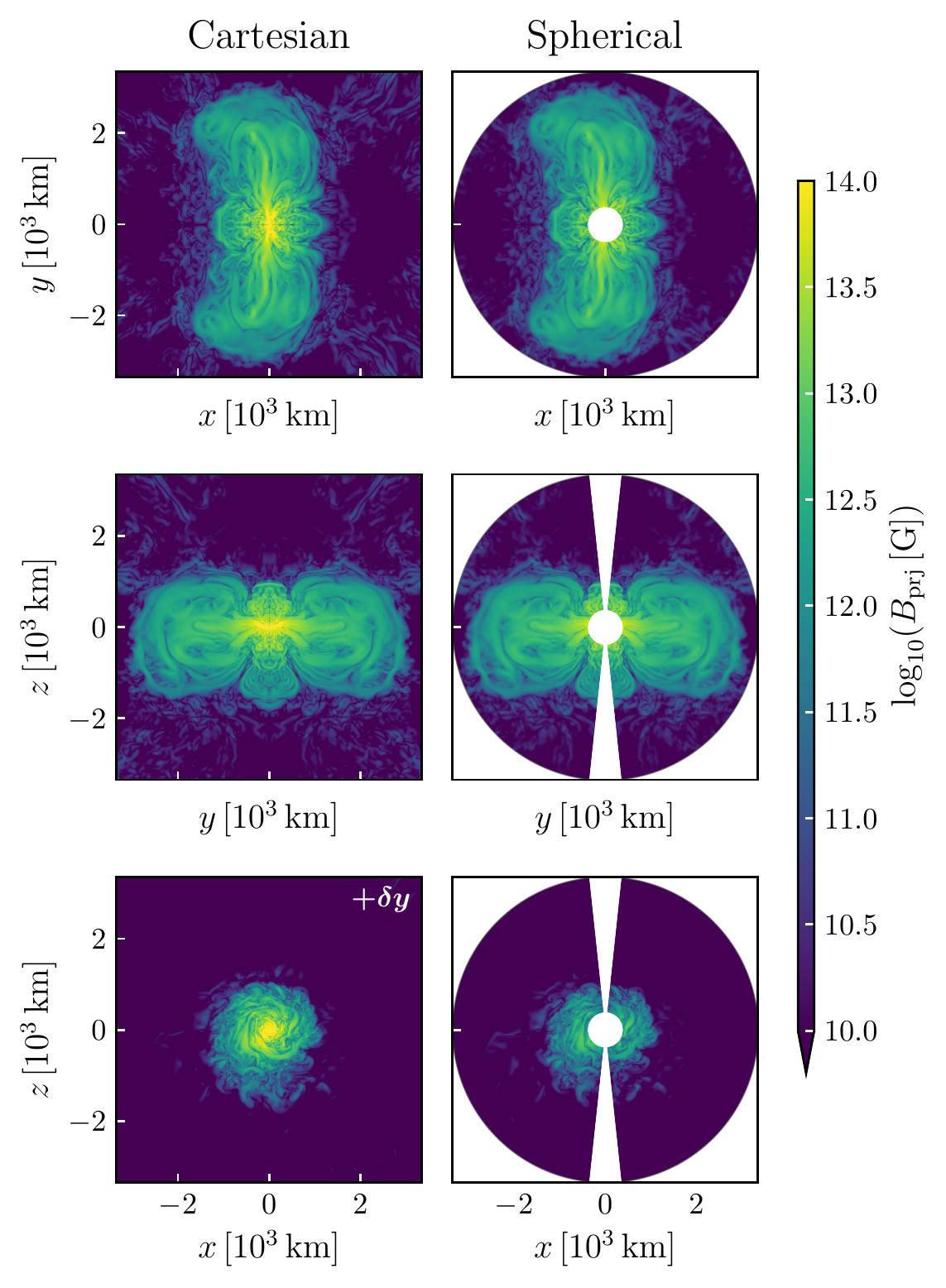}
    \caption{Left: 2D pseudocolor plots of the norm of the projected magnetic field components onto the $xy$, $yz$, and $xz$ planes, after remapping the original data onto the uniform Cartesian grid and applying the $90^{\circ}$ tilting. On the bottom left panel, the plane shown is $y\!=\!\delta y$ (with $\delta y$ the grid spacing along $y$) instead of $y\!=\!0$, which would show zero projected field due to the equatorial reflection symmetry (see text). 
    Right: Corresponding result on the final \textsc{pluto}'s spherical grid. The white regions beyond $\mathrm{\simeq\!3400\,km}$ radius contain the unmagnetized artificial atmosphere employed in our \textsc{pluto} simulations, while the vertical white cone is due to limiting $\theta$ to the $\mathrm{[0.1,\pi-0.1]}$ interval. Finally, the white circle of 380\,km radius corresponds to the excised region not evolved in \textsc{pluto}. See the text for additional details and discussion.}
    \label{fig2}
\end{figure}

To import the magnetic field, we first remap the relative data onto the uniform Cartesian grid, changing the sign of the $x-$ and $y-$field components from $\mathrm{\mathnormal{z}\!>\!0}$ to $\mathrm{\mathnormal{z}\!<\!0}$ to observe the pseudo-vector nature of the field itself. Next, we perform the $90^{\circ}$ tilting, and finally we interpolate all the data onto the \textsc{pluto}'s spherical grid, converting the field components from Cartesian to polar. 
The result of this procedure is illustrated in  Figure\,\ref{fig2}. In particular, the strength of the projected magnetic field is shown for the three fundamental planes, comparing the resulting distribution on the uniform Cartesian grid after $90^{\circ}$ tilting (left) with the final one obtained on the \textsc{pluto}'s spherical grid (right). 
These plots show identical field distributions on the $xy$ and $yz$ planes for the Cartesian and spherical grids. 
The absence of equatorial reflection symmetry gives a non-negligible strength of the projected field on the $xz$ plane for the final spherical grid (see right panel, third row), while for the Cartesian grid the strength would be exactly zero by definition. 
Nevertheless, when considering the Cartesian result on a parallel plane shifted by only one grid cell along the positive $y$-axis (see left panel, third row) the correspondence is excellent, demonstrating that the original field configuration is fully preserved when removing the reflection symmetry.
We further note that, going from the Cartesian to the final spherical grid, no significant variations are observed in the divergence of the magnetic field over the whole 3D domain (in particular, the maximum of the divergence increases only by a factor of $\mathrm{\simeq\!1.6}$).

In Figure\,\ref{fig1} we illustrate some properties of our initial data. In particular, we show 2D meridional views of rest-mass density and radial velocity, where the presence of outflowing material can be appreciated. The mass (outside 380\,km radius) and maximum velocity of such an outflow are $\mathrm{\simeq\!0.17\,\mathnormal{M}_{\odot}}$ and $\mathrm{\simeq\!0.25\,\mathnormal{c}}$, respectively. 
The density distribution is non-uniform and non-isotropic, with an oblate high density inner region surrounded by a lower density and slightly prolate bubble, within which matter accumulates in arc-like structures due to shocks generated during the post-merger phase. The radial velocity profile is non-homologous, with higher values and collimation along the orbital axis of the system, as a result of the activity of the central merger remnant \citep[see][]{Ciolfi2020a}.

\begin{figure*}
	\includegraphics[width=2\columnwidth,keepaspectratio]{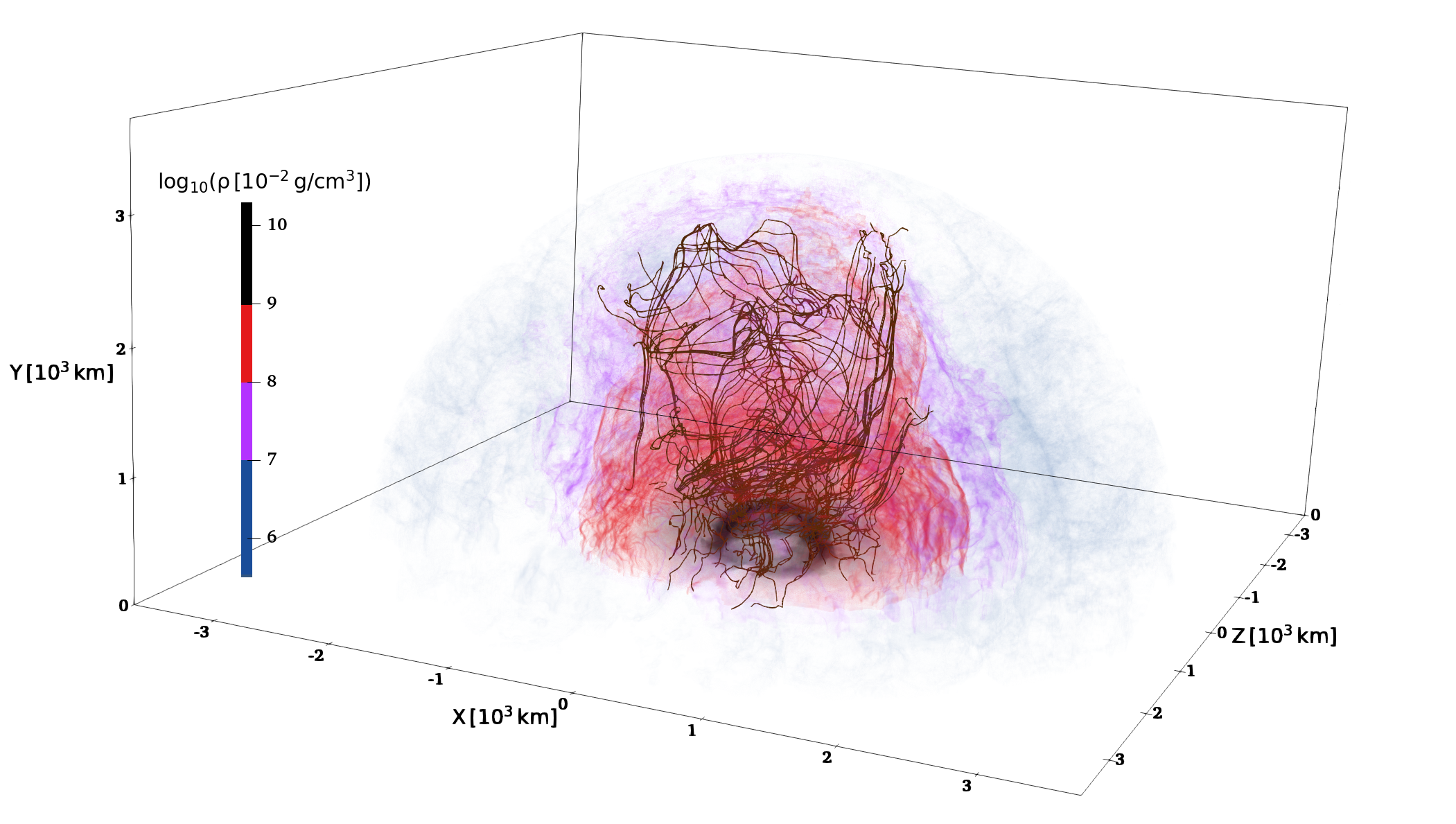}
    \caption{3D rendering of rest-mass density and magnetic field lines for data imported at 155\,ms after merger (as in Figure\,\ref{fig1}). Results are shown only at $\mathrm{\mathnormal{y}\!\ge\!0}$, exploiting the equatorial symmetry of the system. A number of density iso-surfaces is depicted, covering the $\mathrm{\approx\!10^4\!-\!10^8\,g/cm^3}$ range. The magnetic field is represented by means of brown streamlines, which are computed only within regions of magnetic field strength higher than $\mathrm{\simeq\!10^{12}\,G}$. }
    \label{fig3}
\end{figure*}

In Figure\,\ref{fig3} we also provide a 3D rendering of the rest-mass density and magnetic field distributions. The magnetic field, in particular, represented through brown streamlines, shows a denser and twisted structure toward the center, where magneto-rotational effects from the central compact remnant are relevant. Away from the latter, instead, the poloidal-to-toroidal magnetic field ratio increases and the field lines follow the spatial distribution of the ejecta, also accumulating on arc-like structures.

\subsection{Collapse phase}
\label{collapse}

Within the adopted paradigm, the incipient sGRB jet is launched by the BH-disk system resulting from the eventual collapse of the remnant NS. In the intermediate phase between time of collapse ($\mathrm{\mathnormal{t}_c}$) and jet launching, the BH captures material from its surroundings and forms the accretion disk under the combined action of gravitational pull, pressure gradients, and centrifugal support, while the launching mechanism gets activated.

As in P21, we account for the effects of such a pre-jet phase on the environment before inserting the jet itself, using the following prescription:

(i) We introduce a fading acceleration term in the equation of motion that accounts for the gradual removal of the radial pressure gradients caused by BH formation and the consequent accretion of surrounding material. We set this term with the form of an isotropic gravitational `repulsion',
\begin{equation}\label{push}
    \Vec{a} =G\dfrac{M_\mathrm{eff}(r,t)}{r^2}\hat{r} \, , 
\end{equation}
where $G$ is the gravitational constant, and $\mathrm{\mathnormal{M}_{eff}}$ is an `effective mass' that depends on both radial coordinate and time. 
In this way, the extra acceleration manifests itself as a time and radius dependent modification of the central mass responsible for the gravitational pull, i.e.~$M_0-M_\mathrm{eff}(r,t)$ instead of $M_0$.\footnote{We remark that, as in P21, Newtonian gravity is added at all times within the point-mass approximation, setting the source mass as $M_0$, the same of the massive NS remnant (see above).}

(ii) We compute the value of $M_\mathrm{eff}$ at $\mathrm{\mathnormal{r}\!=\!\mathnormal{r}_{exc}}$ and $\mathrm{\mathnormal{t}\!=\!\mathnormal{t}_c}$ by solving the magnetized angle-averaged radial momentum balance equation, in the non-relativistic limit (appropriate for the environment), i.e., 
\begin{equation}\label{MHE}
    M_\mathrm{eff}(r_\mathrm{exc},t_\mathrm{c})
    =-\,\left[r^2\dfrac{1}{\overline{\rho}G}\left(\dfrac{d\overline{P}}{dr}+\dfrac{1}{8\pi}\dfrac{d\overline{B^2}}{dr}-\dfrac{\overline{(\vec{B}\cdot\vec{\nabla})B_r}}{4\pi}\right)\right]_{\mathrm{\mathnormal{r}_{exc},\mathnormal{t}_c}} \, .
\end{equation}

(iii) We shape the function $M_\mathrm{eff}(r,t)$ with linear radial decrease, such that $M_{\mathrm{eff}}$ becomes zero at $\approx\!700$\,km (i.e., roughly twice the excision radius), and with exponential time decay of timescale $\mathrm{\mathnormal{\tau}\!=\!\mathnormal{\tau}_j-(\mathnormal{\tau}_d-\mathnormal{\tau}_j)\sin^2{\mathnormal{\alpha}}}$, where $\mathrm{\mathnormal{\tau}_d}$ is the accretion timescale of the BH-disk system, $\mathrm{\mathnormal{\tau}_j}$ is the delay time between NS collapse and jet launch, and $\alpha$ is the angle with respect to the orbital axis of the system.\footnote{The dependence on $\alpha$ allows us to account for the presence of centrifugal support during accretion (see P21).}
In particular, we set $\mathrm{\mathnormal{\tau}_d\!=\!300\,ms}$ and $\mathrm{\mathnormal{\tau}_j\!=\!30\,ms}$, consistent with BNS merger simulations with incipient jet formation from a BH-disk system \citep[e.g.,][]{Sun2022}.

\subsection{Jet injection}
\label{jet}

At the end of the above phase (i.e., $\mathrm{\mathnormal{\tau}_j\!=\!30\,ms}$ after the NS collapse), we manually inject an incipient sGRB jet into our system.  This is done by means of appropriate radial boundary conditions at $\mathrm{\mathnormal{r}\!=\!\mathnormal{r}_{exc}}$, which we derive using a prescription similar to \cite{Geng2019} \citep[see also][]{Marti2015}. Specifically, we inject a relativistic, magnetized and axisymmetric incipient jet, with uniform rotation and transverse balance between total pressure gradient, centrifugal force, and magnetic tension. Moreover, such a jet has a luminosity that decays over time, with the same timescale as the disk accretion process that is supposed to feed the jet itself (i.e., $\mathrm{\mathnormal{\tau}_d\!=\!300\,ms}$). 

At the time of jet injection, we set the half-opening angle of the jet as $\mathrm{\mathnormal{\theta}_{j}\!=\!10^{\circ}}$, and the initial and terminal Lorentz factors as $\Gamma\mathrm{_{j}\!=\!3}$ and $\Gamma\mathrm{_{\infty}\!=\!300}$, respectively.\footnote{Outside the jet opening angle, we set outflow radial boundary conditions for all physical quantities (including magnetic fields) except for radial velocity, which, as in P21, obeys the same condition as long as $\mathrm{\mathnormal{v^r}\!\le\!0}$ (allowing free falling material to cross the inner radial boundary), otherwise we set $v^r\!=\!0$.}
We start by considering an injection axis parallel to the coordinate $z$-axis, allowing us to make all the primary variables of the jet independent of the azimuthal angle $\phi$ (the $90^{\circ}$ tilting is applied afterwards).
We specify the jet properties through six functions of the polar angle $\theta$ at $r\!=\!r_\mathrm{exc}$, namely, the co-moving density and pressure, $\rho(\theta)$ and $P(\theta)$, and the radial and azimuthal components of 3-velocity and magnetic field, i.e., $v^{r}(\theta),v^{\phi}(\theta),B^{r}(\theta),\textnormal{ and }B^{\phi}(\theta)$.\footnote{We set $v^{\theta}\!=\!B^{\theta}\!=\!0$ as in \cite{Geng2019}.}
To begin with, we adopt a uniform radial velocity profile, such as
\begin{equation}\label{vr}
    v^r=c\sqrt{1-\dfrac{1}{\Gamma^2_{\mathrm{j}}}} \, 
\end{equation}
and, given the assumed uniform rotation of the jet, we set  
\begin{equation}\label{vphi}
    v^{\phi}(\theta)=\overline{\Omega}\,r_\mathrm{exc}\sin{\theta} \, ,
\end{equation}
where $\overline{\Omega}$ is computed by averaging the angular velocity of the surrounding environment, measured at the jet edges in the $yz$-plane. Specifically, we estimate $\mathrm{\overline{\Omega}\!\sim\!10\,rad/s}$ in our fiducial \textsc{pluto} simulation (Section\,\ref{fiducial}). This means a maximum azimuthal velocity of the jet $\mathrm{\sim\!10^{-3}\,c}\!\ll\!v^r$, in accordance with the assumption that $\Gamma_{v^r}\!\simeq\!\Gamma_{\mathrm{j}}$ as adopted in Eq.\,(\ref{vr}).
For the magnetic field, we adopt the same profiles as \cite{Geng2019}, i.e.,
\begin{equation}
    \begin{cases}
        B^{\phi}=\dfrac{2B^{\phi}_{\mathrm{j,m}}(\theta/\theta_{\mathrm{j,m}})}{1+(\theta/\theta_{\mathrm{j,m}})^2} \\
        B^r=B_{\mathrm{ratio}}B^{\phi}_{\mathrm{j,m}} \label{BphiBr}
    \end{cases}\, ,
\end{equation}
with $B^{\phi}_{\mathrm{j,m}}\!=\!6.2\times10^{12}$\,G, $\theta_{\mathrm{j,m}}\!=\!0.4\,\theta_{\mathrm{j}}$, and $B_{\mathrm{ratio}}\!=\!0.5$.
Finally, we compute the co-moving density and pressure of the jet by solving the transverse balance equation coupled with the `two-sided' luminosity of the jet (Appendix\,\ref{transv_eq})
\begin{align}\label{jet_lum}
\begin{split}
    L_{\mathrm{j}}&=4\pi r^2_\mathrm{exc}\int_0^{\theta_\mathrm{j}}\left[\rho h^*\Gamma_{\mathrm{j}}^2c^2-\left(P+\dfrac{b^2}{2}\right)-(b^0)^2\right]v^r\sin{\theta}d\theta \, ,\\
\end{split}
\end{align}
which we set as $L_\mathrm{j}\!=\!3\times 10^{51}$\,erg/s.

In the above expression,  
\begin{equation}\label{h^*}
    h^*=h+\dfrac{b^2}{\rho c^2}=\dfrac{\Gamma_{\mathrm{\infty}}}{\Gamma_{\mathrm{j}}}  
\end{equation}
is the sum of co-moving specific enthalpy ($h$) and ratio of co-moving magnetic energy density 
\begin{equation}\label{b2}
    b^2=\dfrac{1}{4\pi}\left[\dfrac{B^2}{\Gamma_{\mathrm{j}}^2}+\dfrac{(\vec{v}\!\cdot\!\vec{B})^2}{c^2}\right] \, 
\end{equation}
over rest-mass energy density ($\rho c^2$), and
\begin{equation}\label{b0}
    b^0=\dfrac{1}{\sqrt{4\pi}}\Gamma_{\mathrm{j}}(\vec{v}\!\cdot\!\vec{B})/c \, 
\end{equation}
is the time component of the co-moving magnetic field tensor $b^{\mu}$, with $b^2\!=\!b^{\mu}b_{\mu}$. The jet co-moving pressure is given by
\begin{equation}
    P=\dfrac{1}{4}\left[\rho c^2(h^*-1)-b^2\right] \, , \label{jet_prs}
\end{equation}
where $\Gamma_{\mathrm{ad}}\!=\!4/3$ is assumed (appropriate for the incipient jet).

The main jet injection parameters adopted in our simulations are summarized in Table~\ref{inj_param}.
This prescription leads to an initial jet magnetization of
\begin{equation}
    \dfrac{L_{\mathrm{j}}-L_{\mathrm{j,HD}}}{L_{\mathrm{j}}}\simeq8\%
\end{equation}
where $L_{\mathrm{j,HD}}$ corresponds to the purely hydrodynamic luminosity of the jet, computed from Eq.\,(\ref{jet_lum}) with magnetic fields set to zero. Such a magnetization represents a benchmark for all our simulations, and it is directly related to the chosen values of $B^{\phi}_{\mathrm{j,m}}$, $\theta_{\mathrm{j,m}}$, and $B_{\mathrm{ratio}}$.
\begin{table}
    \centering
    \caption{Jet injection parameters: $L_\mathrm{j}$, $\theta_\mathrm{j}$, and $\Gamma_\mathrm{j}$ are the initial luminosity, half-opening angle, and Lorentz factor, respectively; $\Gamma\mathrm{_{\infty}}$ is the terminal Lorentz factor; $\mathrm{\mathnormal{B}_{ratio}}$ is the ratio between maximum toroidal magnetic field strength ($B^{\mathnormal{\phi}}_\mathrm{{j,m}}$ at $\mathrm{\mathnormal{\theta}\!=\!\mathnormal{\theta}_{j,m}}$) and radial magnetic field strength.}
    \begin{tabular}{c|c|c|c|c|c|c|c}
    \hline
    \hline
    $L\mathrm{_{j}}$ &  $\mathrm{\mathnormal{\theta}_{j}}$ & $\Gamma\mathrm{_{j}}$ & $\Gamma\mathrm{_{\infty}}$ & $\mathrm{\mathnormal{B}_{ratio}}$ & $B^{\mathnormal{\phi}}_\mathrm{{j,m}}$ & $\mathrm{\mathnormal{\theta}_{j,m}}$\\
    $\mathrm{[erg/s]}$ & $\mathrm{[^{\circ}]}$ & & & & $\mathrm{[G]}$ & $\mathrm{[^{\circ}]}$ \\
    \hline
    $\mathrm{3\!\times\!10^{51}}$ & $\mathrm{10}$ & $\mathrm{3}$ & $\mathrm{300}$ & $\mathrm{0.5}$ & $\mathrm{6.2\!\times\!10^{12}}$ & $\mathrm{4}$\\
    \hline
    \hline
    \end{tabular}
    \label{inj_param}
\end{table}

To include a time dependence of the jet, we then follow the prescription detailed in Appendix\,\ref{time_pres}. Specifically, we assume $\rho$ and $v^{\phi}$ as constants, and set the time variation of the remaining jet variables as %
\begin{align}
    &v^r(t) = \dfrac{\Gamma_{\mathrm{j}}v^re^{-t/2\tau_{\mathrm{d}}}}{\Gamma(t)} \, ,\label{vr_t}\\ 
    &B^{\phi}(t) = B^{\phi}\sqrt{\dfrac{v^r}{v^r(t)}}e^{-t/2\tau_{\mathrm{d}}} \, ,\label{Bphi_t}\\ 
    &B^{r}(t) = B^{r}\sqrt{\dfrac{v^r}{v^r(t)}}e^{-t/2\tau_{\mathrm{d}}} \, ,\label{Br_t}\\ 
    &P(t) = \dfrac{1}{4}\left[\rho c^2(h^*(t)-1)-b^2(t)\right]\, , \label{prs_t}
\end{align}
where
\begin{align}
    \Gamma(t)=\sqrt{1+\left(\Gamma_{\mathrm{j}}v^re^{-t/2\tau_{\mathrm{d}}}/c\right)^2} \, , \label{Gm_t}
\end{align}
and
\begin{equation}
    h^*(t)=\dfrac{h^*\Gamma_{\mathrm{j}}e^{-t/2\tau_{\mathrm{d}}}}{\Gamma(t)} \, . \label{h*_t}
\end{equation}
When combined together, these profiles lead to an injection luminosity that varies as $L(t)\!\simeq\!L_{\mathrm{j}}e^{-t/\tau_{\mathrm{d}}}$.
 
As final step, we apply the $90^{\circ}$ tilting to obtain the jet prescription to be adopted in our simulations (with injection axis orthogonal to the coordinate polar axis). This is detailed in Appendix\,\ref{vec_tilt}.

\section{Extending the explorable range of jet launching times}
\label{extr}

The (maximum) data import time of our \textsc{pluto} simulations, namely, $\mathrm{\mathnormal{t}_0\!\simeq\!155\,ms}$ after merger, is constrained by the spatial domain boundaries of the reference BNS merger simulation from which data are taken (see Figure\,\ref{fig1} and  Section\,\ref{import}). Such time, however, represents only a fraction of the total time span covered by the reference simulation itself, extending up to $\mathrm{\simeq\!255\,ms}$ after merger. Therefore, importing data around 155\,ms leaves about 100\,ms of further remnant evolution unexploited. 
Moreover, the inability to access later times in the life of the remnant NS forces us to consider collapse times $t_\mathrm{c}\!\le\!155$\,ms (see Section\,\ref{collapse}), 
while according to numerous investigations on GW170817/GRB\,170817A, the metastable NS resulting from the coalescence survived much longer, from a few hundred ms to $\sim$1\,s \citep[e.g.,][]{Metzger2018,Gill2019,Zhang2019,Lazzati2020,Beniamini2020,Nakar2020,Murguia2021}. 

In light of the above limitations, we discuss in the following a procedure we developed to (i) import information from later post-merger times, fully exploiting the reference BNS merger simulation, and (ii) consider collapse times (and hence, jet launching times) significantly later than 155\,ms after merger, more compatible with GW170817/GRB\,170817A. 
In this work, we consider in particular a fiducial collapse time of $\simeq\!355$\,ms after merger (see Section\,\ref{fiducial}), which would not be possible without such a procedure.

\subsection{Time dependent boundary conditions at \texorpdfstring{$\boldsymbol{\mathrm{\mathnormal{r}\!=\!\mathnormal{r}_{exc}}}$}{rexc}}\label{tdepen}

\begin{figure}
	\includegraphics[width=\columnwidth,keepaspectratio]{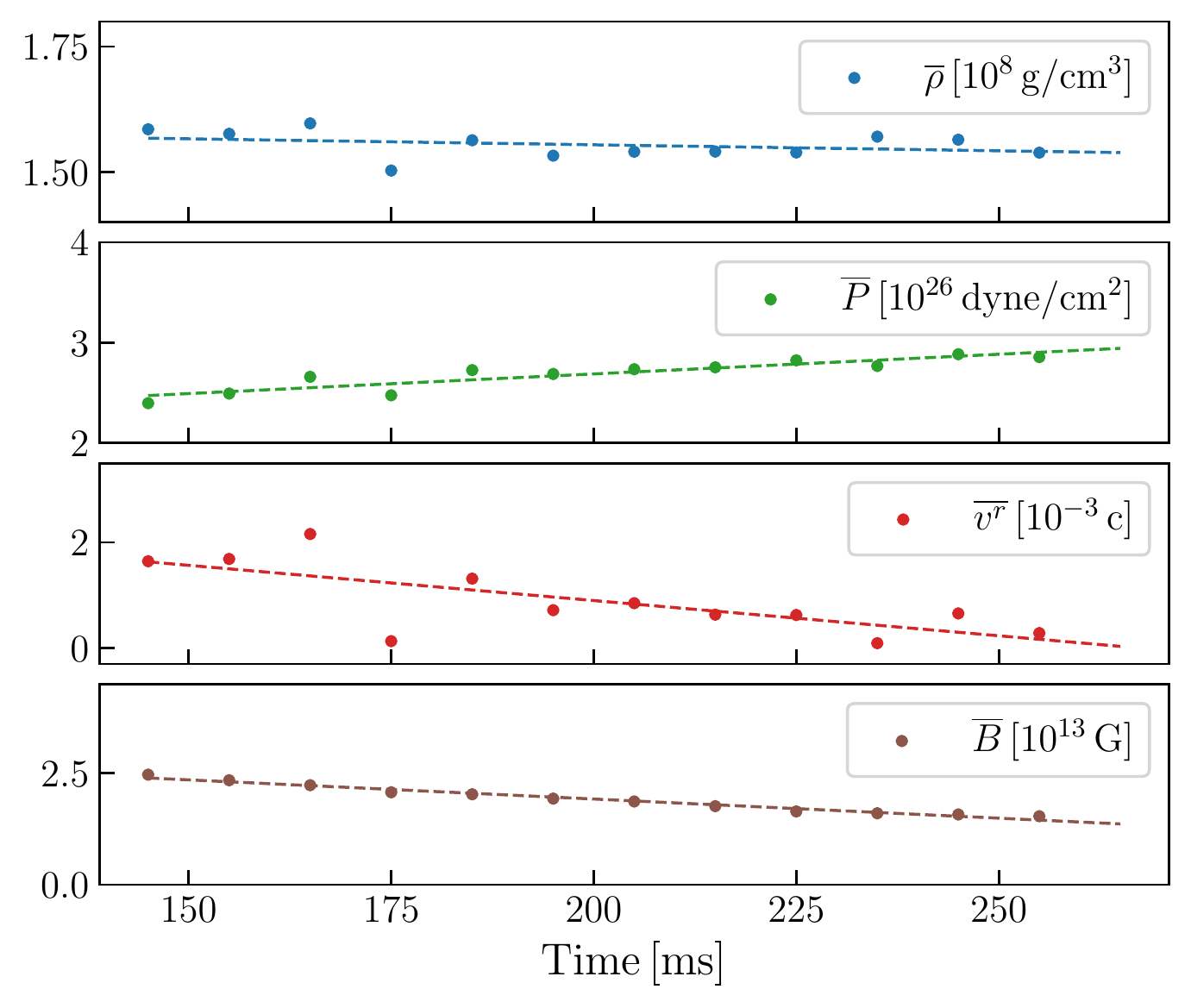}
    \caption{Top to bottom: Time evolution of angle-averaged rest-mass density, pressure, radial velocity, and magnetic field strength at $\mathrm{\mathnormal{r}\!=\!\mathnormal{r}_{exc}\,(=\!380\,km)}$. The colored dots indicate values measured from the reference BNS merger simulation at 10\,ms steps from 145\,ms after merger. The colored dashed lines are linear fits to the measured values. See text for additional details.}
    \label{fig4}
\end{figure}

In order to consider collapse times later than the time of data import, $\mathrm{\mathnormal{t}_0\!\simeq\!155\,ms}$ after merger, we need to further evolve the system in the pre-collapse phase using our \textsc{pluto} setup, reproducing as close as possible the evolution given by the original general relativistic simulation with no central excision.
As already shown in P21, a fairly accurate dynamics can be obtained by setting appropriate time-dependent radial boundary conditions at $\mathrm{\mathnormal{r}\!=\!\mathnormal{r}_{exc}\,(=\!380\,km)}$, according to the following steps:

(i) We first compute (at $r_\mathrm{exc}$) the linear fits to the original time evolution of the angle-averaged rest-mass density, pressure, radial velocity, and magnetic field strength (based on the full BNS merger simulation time span, in this case up to $\mathrm{\simeq\!255\,ms}$ after merger; see Figure\,\ref{fig4}).

(ii) We then multiply the spatial distribution of each primary fluid variable at $\mathrm{\mathnormal{r}\!=\!\mathnormal{r}_{exc}}$ and $\mathrm{\mathnormal{t}\!=\!\mathnormal{t}_0}$ by the corresponding time evolution trend obtained in (i).\footnote{For example, the rest-mass density distribution $\rho(r_\mathrm{exc},\theta,\phi,t_0)$ is multiplied by the function of time $\overline{\rho}(r_\mathrm{exc},t)/\overline{\rho}(r_\mathrm{exc},t_0)$, where $\overline{\phantom{x}}$ means angle-averaged and $\overline{\rho}(r_\mathrm{exc},t)$ is the result of the linear fit.} 
For $B^r$, $B^{\theta}$, and $B^{\phi}$ we assume the same trend as the full magnetic field strength, and we set constant $v^{\theta}$ and $v^{\phi}$ as in P21.

(iii) We finally adopt the time-dependent distributions resulting from (ii) as radial boundary conditions at $\mathrm{\mathnormal{r}\!=\!\mathnormal{r}_{exc}}$.

In Figure\,\ref{fig4}, we show the original evolution of angle-averaged rest-mass density, pressure, radial velocity, and magnetic field strength at $\mathrm{\mathnormal{r}\!=\!\mathnormal{r}_{exc}}$, from 145\,ms after merger.
Overall, data from the reference BNS merger simulation reveal a decrease in rest-mass density, radial velocity, and magnetic field strength, while pressure increases. 
We also note that linear fits describe rather well the evolution, in particular from $\mathrm{\simeq\!185\,ms}$ after merger, indicating a quite regular behaviour of the system at that stage.
For this reason, we can also consider using the same linear fits to extrapolate the trends beyond the last point from the original merger simulation (at $\mathrm{\simeq\!255\,ms}$ after merger). 
In this work, we extend the pre-collapse phase by another 100\,ms, up to $\mathrm{\simeq\!355\,ms}$ after merger, assuming that our fits are still representative (see also the discussion at the end of the next Section). 

We note that, following the linear fit, the average radial velocity at $\mathrm{\mathnormal{r}\!=\!\mathnormal{r}_{exc}}$ becomes negative around $\mathrm{\simeq\!275\,ms}$ after merger, in agreement with the expectation that bound matter around the remnant NS would eventually start to fall back \citep{CiolfiKalinani2020}, 
while the average magnetic field strength remains always positive, as it should, for the times considered (with this linear trend, it would change sign around $\mathrm{\simeq\!420\,ms}$).

\subsection{Substitution steps}
\label{substitution} 

The approach described in the previous Section can be further improved via additional `substitution' steps, in which we re-import, after a time interval $\Delta t_\mathrm{sub}$, the full 3D data of rest-mass density, pressure, 3-velocity, and magnetic field from the reference BNS merger simulation up to a certain radius $\mathrm{\mathnormal{r}_{sub}\!=\!\mathnormal{v}_{max}\Delta\mathnormal{t}_{sub}}$, where $\mathrm{\mathnormal{v}_{max}}$ is the maximum radial velocity of the original system at $\mathrm{\mathnormal{r}\!=\!\mathnormal{r}_{exc}}$ during the considered evolution time span.
As a result, when applying such a substitution we remove any discrepancy with the original merger simulation up to the radius ($\mathrm{\mathnormal{r}_{sub}}$) below which the system is dynamically connected with the ongoing evolution at $\mathrm{\mathnormal{r}_{exc}}$ (for the given time interval $\Delta t_\mathrm{sub}$). Outside this substitution radius, the system is essentially unaffected by differences in the evolution at $\mathrm{\mathnormal{r}_{exc}}$, and thus no modification is required.

Specifically, we estimate $\mathrm{\mathnormal{v}_{max}\!\simeq\!0.12}\,c$ and adopt $\Delta t_\mathrm{sub}\!=\!50$\,ms, yielding $\mathrm{\mathnormal{r}_{sub}\!\simeq\!1800\,km}$. Data substitution is first performed at $\mathrm{\simeq\!205\,ms}$ after merger and another time at $\mathrm{\simeq\!255\,ms}$ after merger, thus taking full advantage of the results provided by the general-relativistic simulation.
\begin{figure}
	\includegraphics[width=\columnwidth,keepaspectratio]{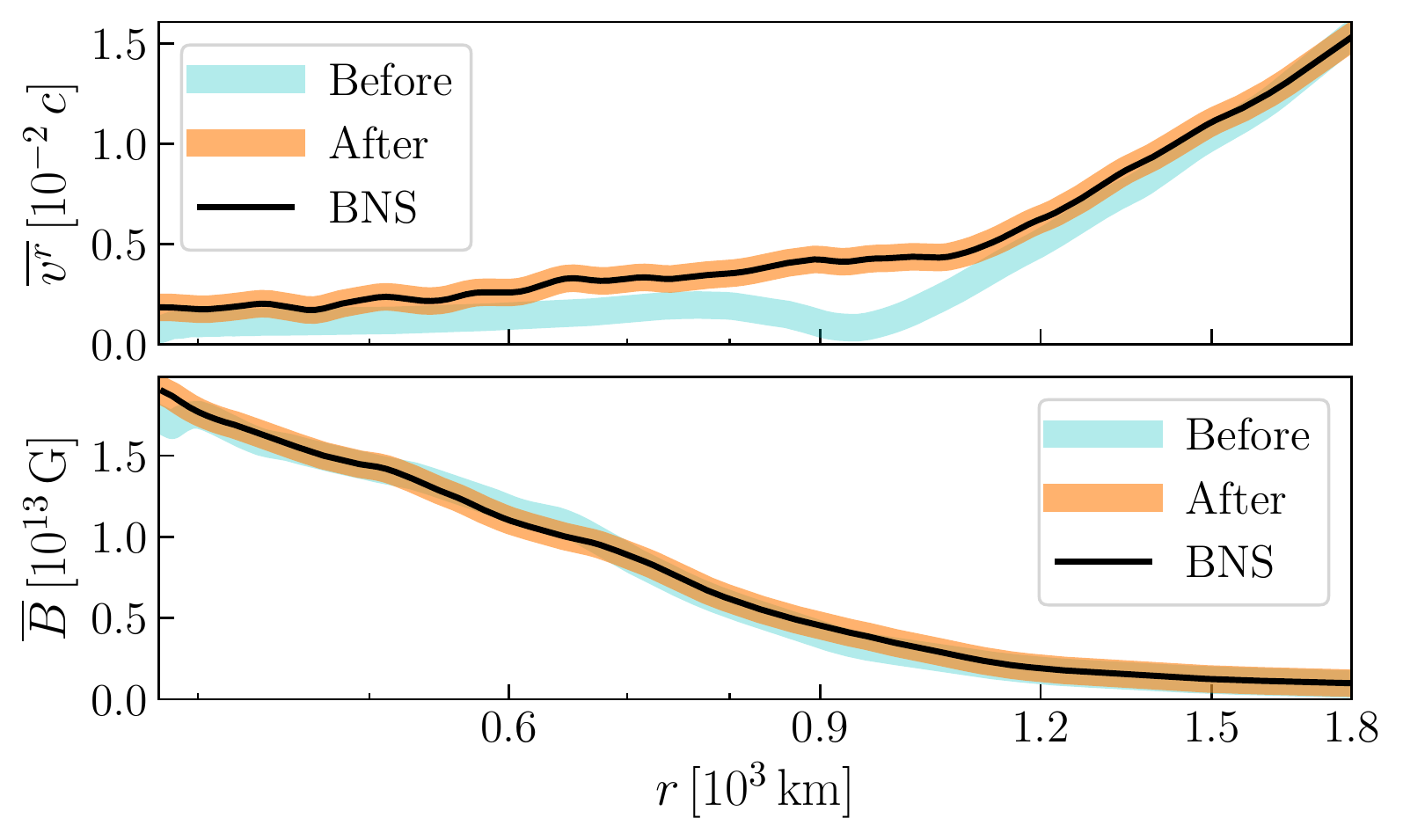}
    \caption{Radial profiles of angle-averaged radial velocity (top) and magnetic field strength (bottom) at $\mathrm{\simeq\!205\,ms}$ after merger. Thick-colored lines indicate the profiles obtained in \textsc{pluto} before (cyan) and after (orange) the substitution step (see text). Solid-black lines show instead the same profiles extracted from the reference BNS merger simulation at the same post-merger time.}
    \label{fig5}
\end{figure}

In Figure\,\ref{fig5}, we illustrate the effects of our data substitution at $\mathrm{\simeq\!205\,ms}$ after merger. 
In particular, with thick-colored lines we show the radial profiles (up to $r_\mathrm{sub}$) of angle-averaged radial velocity (top) and magnetic field strength (bottom) before and after the substitution itself. Over-plotted as solid-black lines, we also show the profiles from the BNS merger simulation at the same post-merger time. 
As shown in the Figure, after 50\,ms of \textsc{pluto} evolution some differences emerge with respect to the original simulation, in particular in the radial velocity.
Such differences, however, are eventually removed via the substitution step.
\begin{figure*}
	\includegraphics[width=1.8\columnwidth,keepaspectratio]{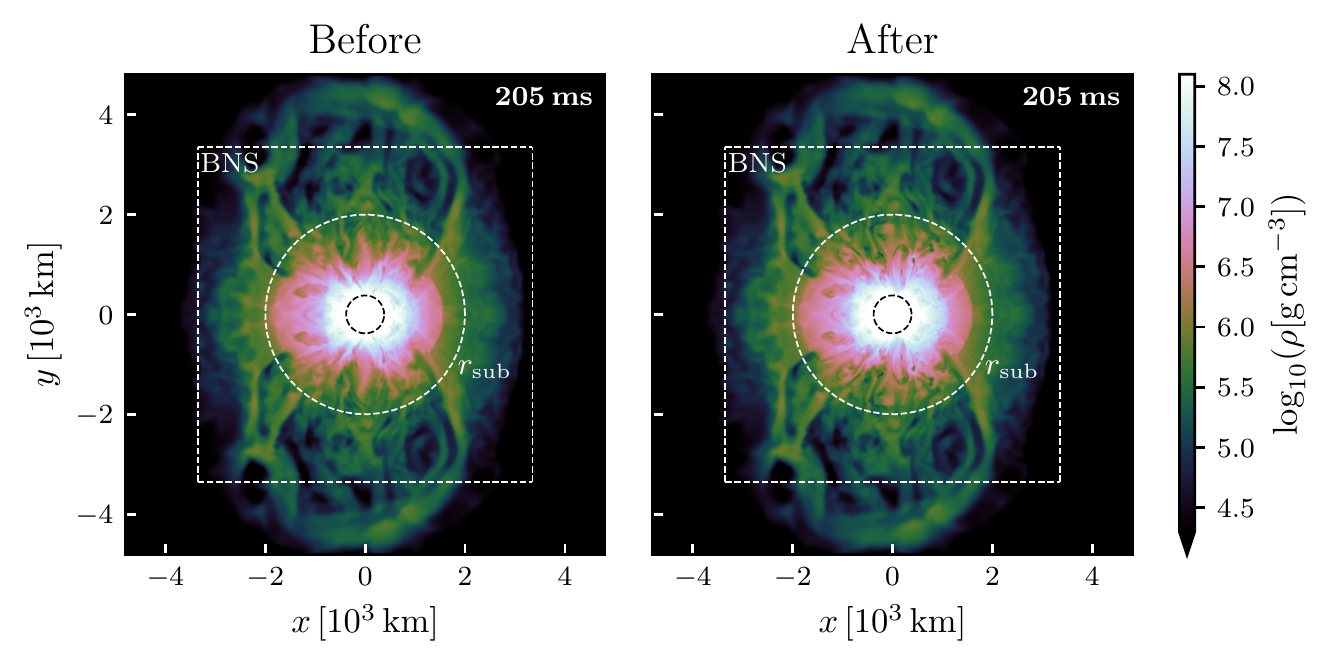}
    \caption{Meridional view of rest-mass density at $\mathrm{\simeq\!205\,ms}$ after merger before (left) and after (right) the substitution step (see text). In both panels, white-dashed contours indicate the computational domain of the reference merger simulation (square labelled `BNS') and the substitution region (circle labelled `$\mathrm{\mathnormal{r}_{sub}}$'). The inner black-dashed circle marks the excision radius $\mathrm{\mathnormal{r}_{exc}\!=\!380\,km}$.}
    \label{fig6}
\end{figure*}

In Figure\,\ref{fig6}, we show 2D snapshots of rest-mass density at $\mathrm{\simeq\!205\,ms}$ after merger, before (left) and after (right) data substitution. In both panels, we also indicate with white-dashed contours the computational domain of the original merger simulation (square labelled `BNS') and the substitution region (circle labelled `$\mathrm{\mathnormal{r}_{sub}}$'). 
While significant changes are present within the substitution region, a smooth transition remains at the radius $\mathrm{\mathnormal{r}_{sub}\!\simeq\!1800\,km}$.
We also remark that the system evolution is self-consistently reproduced even outside the limited domain in which original BNS merger data can be imported. 

In conclusion, the adopted procedure allows us to overcome the aforementioned limitations in employing data imported from the reference BNS merger simulation (see discussion at the beginning of Section\,\ref{extr}). 
At $\mathrm{\simeq\!255\,ms}$ after merger, we obtain a system that reproduces the exact physical quantities of the reference simulation, while extending beyond the original computational domain. Then, we can exploit the high regularity of the late-time evolution (Section\,\ref{tdepen}) to continue, for a reasonable time, beyond the reach of the merger simulation without introducing significant inaccuracies. 
We choose to evolve in \textsc{pluto} for another 100\,ms, up to $\mathrm{\simeq\!355\,ms}$ after merger. 
Given that $\mathrm{\simeq\!255\,ms}$ after merger the maximum radial velocity at $\mathrm{\mathnormal{r}_{exc}}$ is only $\mathrm{\mathnormal{v}_{max}\!\simeq\!0.02}\,c$ and decreasing, the only portion of the system directly influenced by our time-varying boundary conditions over another 100\,ms of evolution is the one between 380 and less than 1000\,km radius.
Moreover, possible deviations with respect to a fully general relativistic evolution with no central excision are expected to be much smaller at these late times.

\section{Fiducial model}
\label{fiducial}

We now turn to discuss the results of our fiducial \textsc{pluto} simulation performed using the setup described in the previous Sections. First, in Section\,\ref{launch}, we focus on the system evolution up to the time of jet injection (namely, $\mathrm{\simeq\!385\,ms}$ after merger). Then, in Section\,\ref{prop}, we discuss the jet breakout and further propagation, providing a quantitative description of the main evolutionary stages.

\subsection{Evolution before jet injection}\label{launch}

From time of data import $t_0$ to $\mathrm{\simeq\!355\,ms}$ after merger, the system is evolved using the numerical approach described in Section\,\ref{extr}. 
Specifically, the material influx at $\mathrm{\mathnormal{r}\!=\!\mathnormal{r}_{exc}}$ is reproduced via the time linear profiles shown in Figure\,\ref{fig4}. Furthermore, the system behaviour within the substitution radius $\mathrm{\mathnormal{r}\!\le\!\mathnormal{r}_{sub}}$ is kept consistent with the original BNS merger simulation by carrying out additional data import and substitution at $\mathrm{\simeq\!205\,ms}$ (see Figure\,\ref{fig6}) and again at $\mathrm{\simeq\!255\,ms}$ after merger, so that the time coverage of the original simulation is exploited in full. 

At $\mathrm{\simeq\!355\,ms}$ after merger, the remnant NS is assumed to collapse and the system evolution is followed for other $\mathrm{30\,ms}$ using the post-collapse prescription described in Section\,\ref{collapse}. In particular, we estimate $\mathrm{\mathnormal{M}_{eff}(\mathnormal{r}_{exc},\mathnormal{t}_{c})\!\simeq\!0.985\,\mathnormal{M}_{\odot}}$ from Eq.\,(\ref{MHE}), which we employ to mimic the aftermath of the BH-disk system formation as discussed in the same Section.

\begin{figure}
	\includegraphics[width=\columnwidth,keepaspectratio]{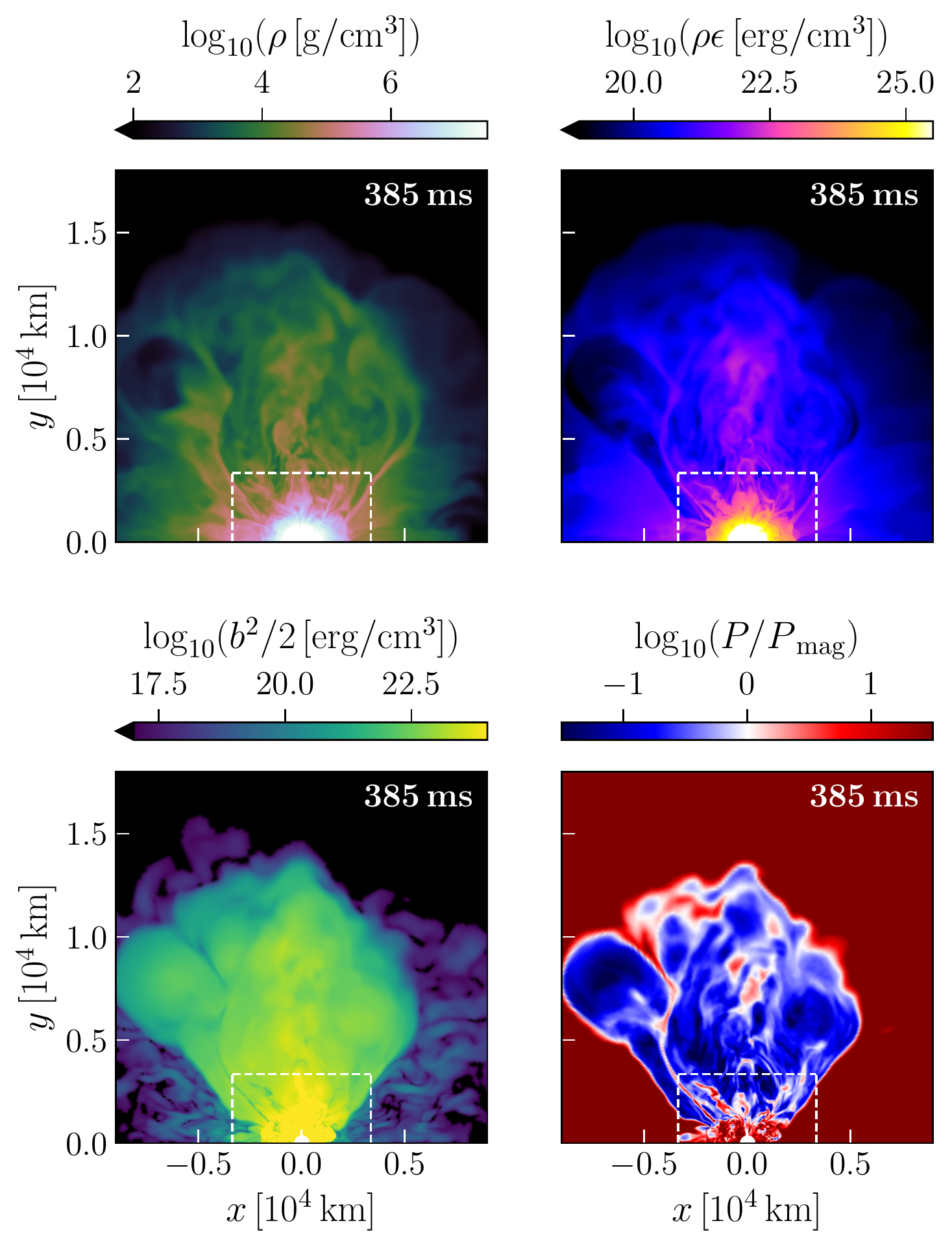}
    \caption{System configuration at $\mathrm{\simeq\!385\,ms}$ after merger (time of jet injection). Each panel shows quantities in the $xy$ plane (for $y\ge 0$). In particular, the upper panels show the distribution of rest-mass density (left) and internal energy density (right), while the lower panels the corresponding magnetic field energy density (left) and the thermal to magnetic pressure ratio (right). 
    As the dashed-white contours indicate, the system extends well beyond the domain boundaries of the reference BNS merger simulation at this stage (see text for further discussion).}
    \label{fig7}
\end{figure}

At $\mathrm{\simeq\!385\,ms}$ after merger (i.e.~the time of jet injection), we achieve the system configuration shown in Figure\,\ref{fig7}. At this time, the merger ejecta extends up to $\mathrm{\simeq\!1.5\!\times\!10^4\,km}$ (well beyond the domain boundaries of the reference BNS merger simulation), and has $\mathrm{\simeq\!30\%}$ more mass than the time of data import. In particular, the rest-mass density distribution (top-left panel) is rather uniform and equatorially symmetric, with a compact region in the center surrounded by a large, less dense petal-shaped bubble. The corresponding internal energy density distribution (top-right panel) peaks toward the most central regions, decreasing by several orders of magnitude at large scales. In contrast, the magnetic field energy density distribution (bottom-left panel) is fairly uniform, even at large scales, causing the ejecta to be magnetically dominated for the most part (see bottom-right panel).

\subsection{Jet breakout and propagation}\label{prop}

The incipient sGRB jet is manually injected into the above system using the prescription detailed in Section\,\ref{jet} (also see Appendix\,\ref{jet_calc}). 
Immediately after that, it attempts to pierce through the surrounding warm, high-density region, dissipating much of its kinetic energy into heat. 
After $\mathrm{\simeq\!100\,ms}$, it successfully pushes away the dense material in front of it, generating a forward shock that diverts sideways the material itself.  
Consequently, a high-pressure cocoon is formed around the jet, keeping the latter collimated during propagation. The collimation leads to the formation of an oblique shock at the jet base, which limits lateral widening, and increases the ram pressure into the jet funnel. As a result, the jet front is pushed to higher velocities and the rate of energy flow into the cocoon is significantly reduced \citep{Bromberg2011}. 

\begin{figure*}
	\includegraphics[width=2\columnwidth,keepaspectratio]{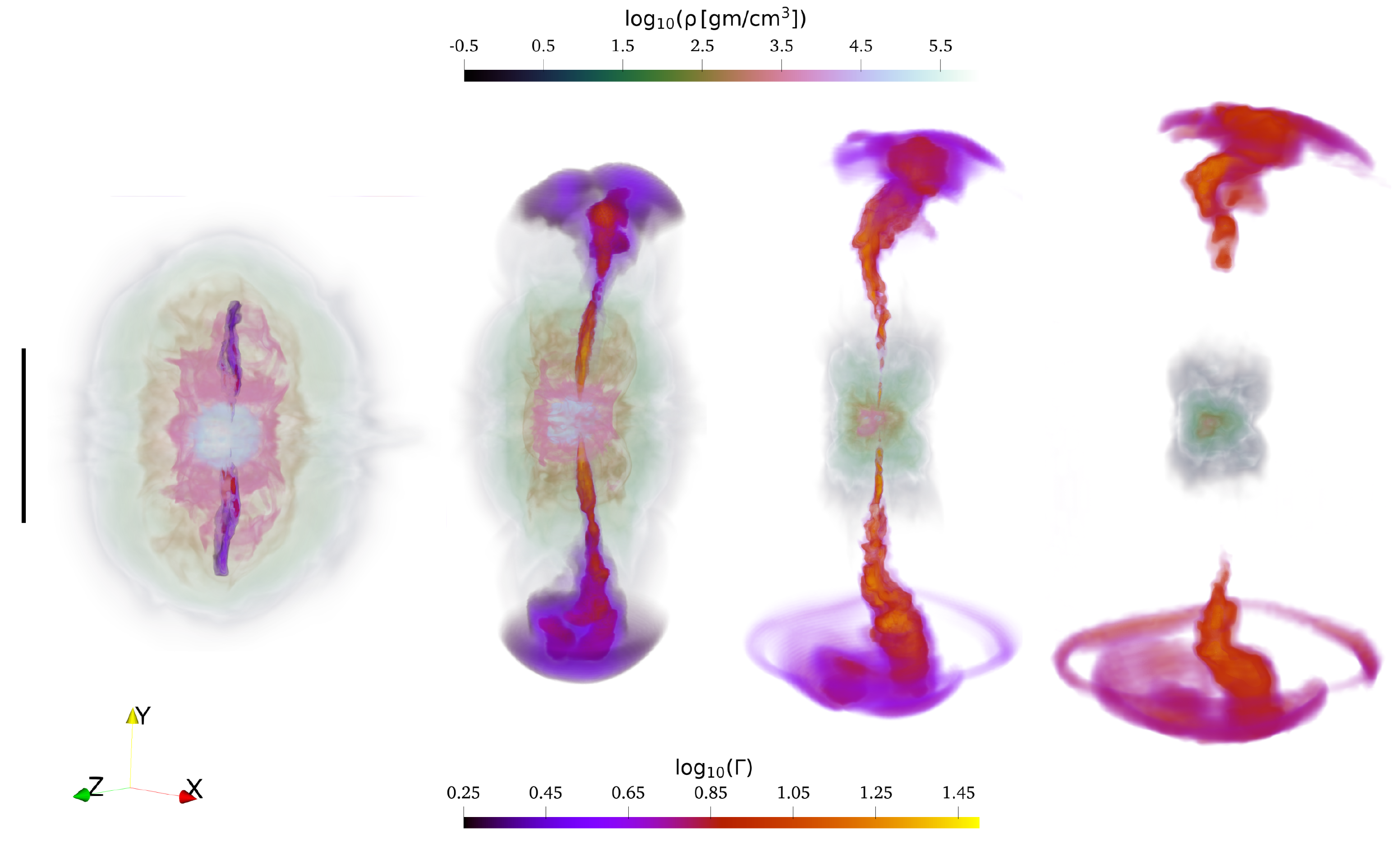}
    \caption{Jet-environment system configuration at four different times after merger. From left to right, in particular, we show 3D renderings of rest-mass density and Lorentz factor at 560\,ms, 735\,ms, 1385\,ms, and 2385\,ms after merger (namely, 175\,ms, 350\,ms, 1000\,ms, and 2000\,ms after jet injection). The black-colored bar corresponds to 0.2, 0.5, 1.8, and 3.6 times $10^5$\,km, respectively. See text for discussion. }
    \label{fig8}
\end{figure*}

To illustrate the above process, we show in the first 3D rendering on the left of Figure\,\ref{fig8} the jet-environment system configuration at 175\,ms after injection (namely, 560\,ms after merger). In particular, shown are the rest-mass density and Lorentz factor distributions for $|y|\!\le\!3\times10^4$\,km.
We notice the presence of a dense material shell around the jet with average density $\mathrm{\simeq\!3\times10^3\,g/cm^3}$ and spatial extension $\mathrm{\simeq\!2\!\times\!10^4\,km}$. Toward the jet base, in particular, such a shell provides high collimation to the jet, allowing the latter to accelerate up to Lorentz factors $\mathrm{\sim\!10}$. At the jet front, the interplay with the realistic environment causes the jet to `snake' along regions of lower density, breaking the axisymmetry adopted at injection.

The jet evolution proceeds similarly as above until the breakout, occurring at $\mathrm{\simeq\!350\,ms}$ after injection (second rendering from the left in Figure\,\ref{fig8}). At this stage, the cocoon still effectively collimates the jet, allowing the latter to accelerate up to $\mathrm{\Gamma\!\simeq\!30}$ along the main propagation axis. We note the high stability of the jet-environment interface up to the breakout radius (namely, $r_\mathrm{b}\!\simeq\!5\!\times\!10^4$\,km). This may be related to the presence of magnetic fields, which prevent the rising of turbulent motions inside the jet cone as those observed in similarly realistic but non-magnetized systems (see P21).
Above such radius, the jet front exhibits an asymmetric profile resulting from previous interactions with the surrounding material. Such a degree of asymmetry in the outflow, observed in presence of a realistic environment, is not found in sGRB jet simulations which adopt hand-made initial conditions for the latter.

At $\mathrm{\simeq\!1\,s}$ after injection, the time decay of the jet injection luminosity (see Appendix\,\ref{time_pres}) begins to take effect. In particular, this causes the jet to break down at the base, gradually detaching from the surrounding environment (third rendering from the left in Figure\,\ref{fig8}). Meanwhile, the jet front accelerates and expands laterally. 

Finally, at $\mathrm{\simeq\!2\,s}$ after injection (rendering on the right), the detachment of the jet is fully achieved, and the jet `head' (namely, the high Lorentz factor portion of the jet) becomes more compact as the front propagates more slowly than the ultra-relativistic tail behind it.

The overall physical system reaches the final configuration shown in Figure\,\ref{fig9}. At this stage, the jet head propagates at maximum Lorentz factor of $\simeq\!40$ (reached slightly outside the $xy$-plane shown in the Figure), extending from $\mathrm{\simeq\!3\!\times\!10^5\,km}$ to $\mathrm{\simeq\!6\!\times\!10^5\,km}$. Separately, the sub-relativistic environment expands up to $\mathrm{\simeq\!4\!\times\!10^5\,km}$, containing most of the mass and the internal energy of the system (first to third panels of Figure\,\ref{fig9}). In particular, the conversion of magnetic energy into kinetic energy after breakout causes the first to remain mostly confined to the environment (fourth panel of Figure\,\ref{fig9}), resulting in a jet head that is kinetically dominated. 

We show the angular properties of the jet head in Figure\,\ref{fig10}, reporting distributions of the radial-averaged Lorentz factor $\mathrm{\overline{\Gamma}}$ (weighted over the total energy density) and isotropic equivalent energy $\mathrm{E_{iso}}$ as a function of polar and azimuthal angles, at both the north and south jet sides. We notice the presence of significant asymmetry of the jet head with respect to both the main propagation axis and the equatorial plane, with maxima $\mathrm{\overline{\Gamma}_{max}\!\simeq\!27.9}$ and $\mathrm{E_{iso,max}\!\simeq\!1.0\times10^{53}\,erg}$ at the north pole, while $\mathrm{\overline{\Gamma}_{max}\!\simeq\!22.1}$ and $\mathrm{E_{iso,max}\!\simeq\!1.3\times10^{53}\,erg}$ at the south pole (dashed-gray crosshairs in the panels). Especially, we point out the lack of Gaussianity, typically assumed in current sGRB jet afterglow models.

\section{Concluding remarks}
\label{summary}

We presented the first 3D-RMHD simulation of an incipient magnetized sGRB jet propagating across the realistic environment surrounding the remnant of a magnetized BNS coalescence. 
Here, by `realistic' we mean that the environment (defined by specific distributions of rest-mass density, pressure, 3-velocity, and magnetic field) was directly imported from the outcome of a general-relativistic magnetized BNS merger simulation (Section\,\ref{import}).

For the case at hand, the incipient jet was launched $\simeq\!385$\,ms after merger, corresponding to 30\,ms after the assumed collapse time of the massive NS remnant into a BH.
The evolution of the environment from the initial data import up to the jet launching time was obtained by applying a newly developed procedure to fully exploit the information from the reference BNS merger simulation, covering up to $\simeq\!255$\,ms after merger, and then further extrapolating up to the chosen collapse time of $\simeq\!355$\,ms after merger (Section\,\ref{extr}). For the intermediate phase between collapse and jet launching, lasting 30\,ms, we evolved the system according to the scheme presented in P21, now generalized to account for magnetic fields as well (Section\,\ref{collapse}).
The incipient jet itself was modelled by means of an advanced and physically motivated prescription that incorporates uniform rotation, magnetic fields, and both temporal and angular dependencies for the relevant physical variables (Section~\ref{jet}).
Finally, we followed the evolution of the jet and the whole system for a time span of $\mathrm{2\,s}$, covering spatial scales in excess of $10^5\,{\rm km}$, and carried out a detailed analysis of the jet-environment interplay process and outcome (Section\,\ref{fiducial}).
\begin{figure*}
	\includegraphics[width=1.9\columnwidth,keepaspectratio]{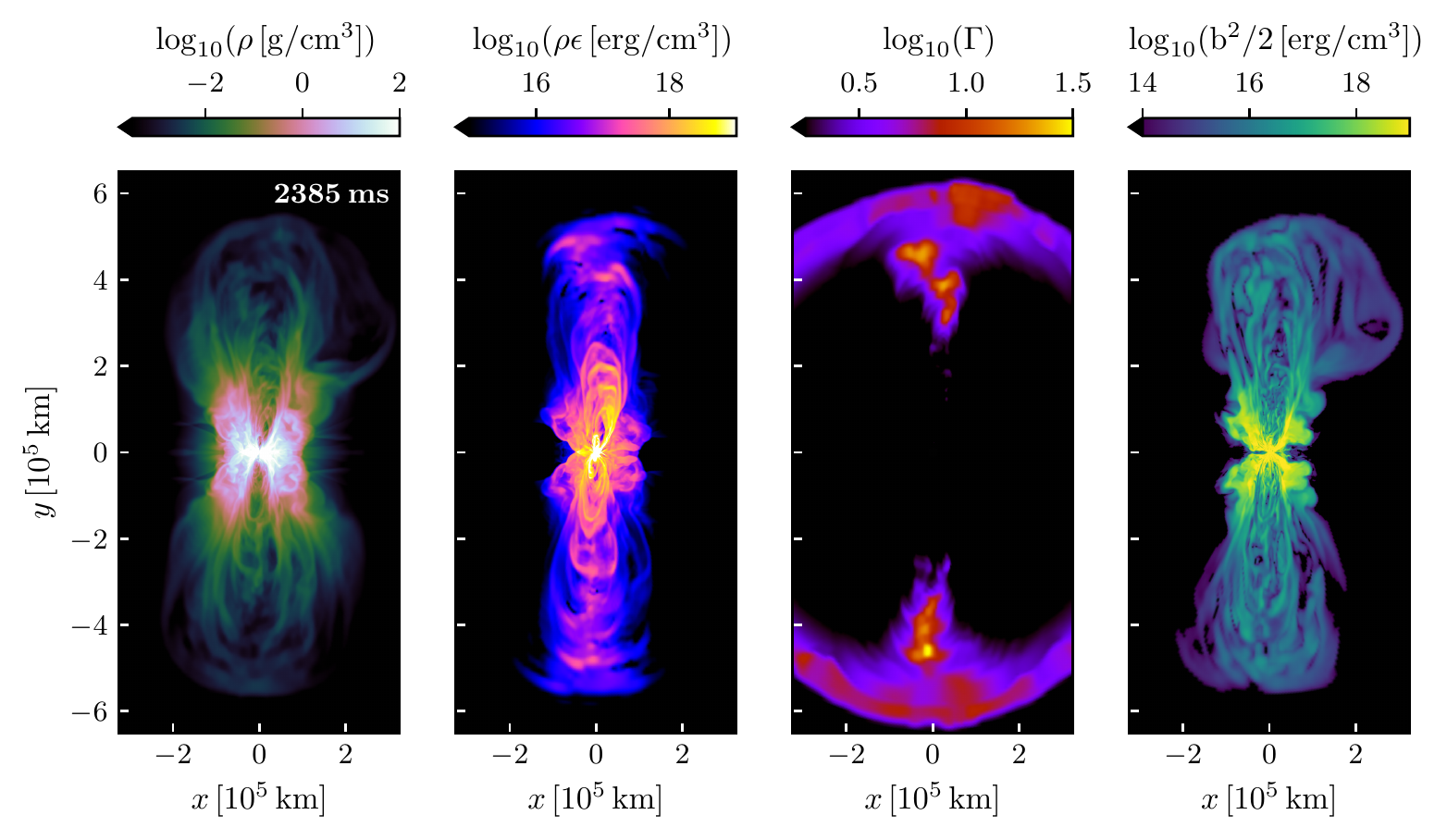}
    \caption{Jet-environment system configuration at $\mathrm{\simeq\!2385\,ms}$ after merger (namely, $\mathrm{\simeq\!2\,s}$ after jet injection). From left to right, we show the meridional distribution of rest-mass density, internal energy density, Lorentz factor, and magnetic field energy density (all densities are computed in the co-moving frame of the jet). We notice the presence of a jet head, extending from $\mathrm{\simeq\!3\!\times\!10^5\,km}$ to $\mathrm{\simeq\!6\!\times\!10^5\,km}$, which propagates at maximum Lorentz factor of $\mathrm{\simeq\!40}$ (reached outside the $xy$-plane). Separately, the sub-relativistic environment expands up to $\mathrm{\simeq\!4\!\times\!10^5\,km}$, containing most of the mass and internal and magnetic energy of the system. }
    \label{fig9}
\end{figure*}
\begin{figure}
    \includegraphics[width=\columnwidth,keepaspectratio]{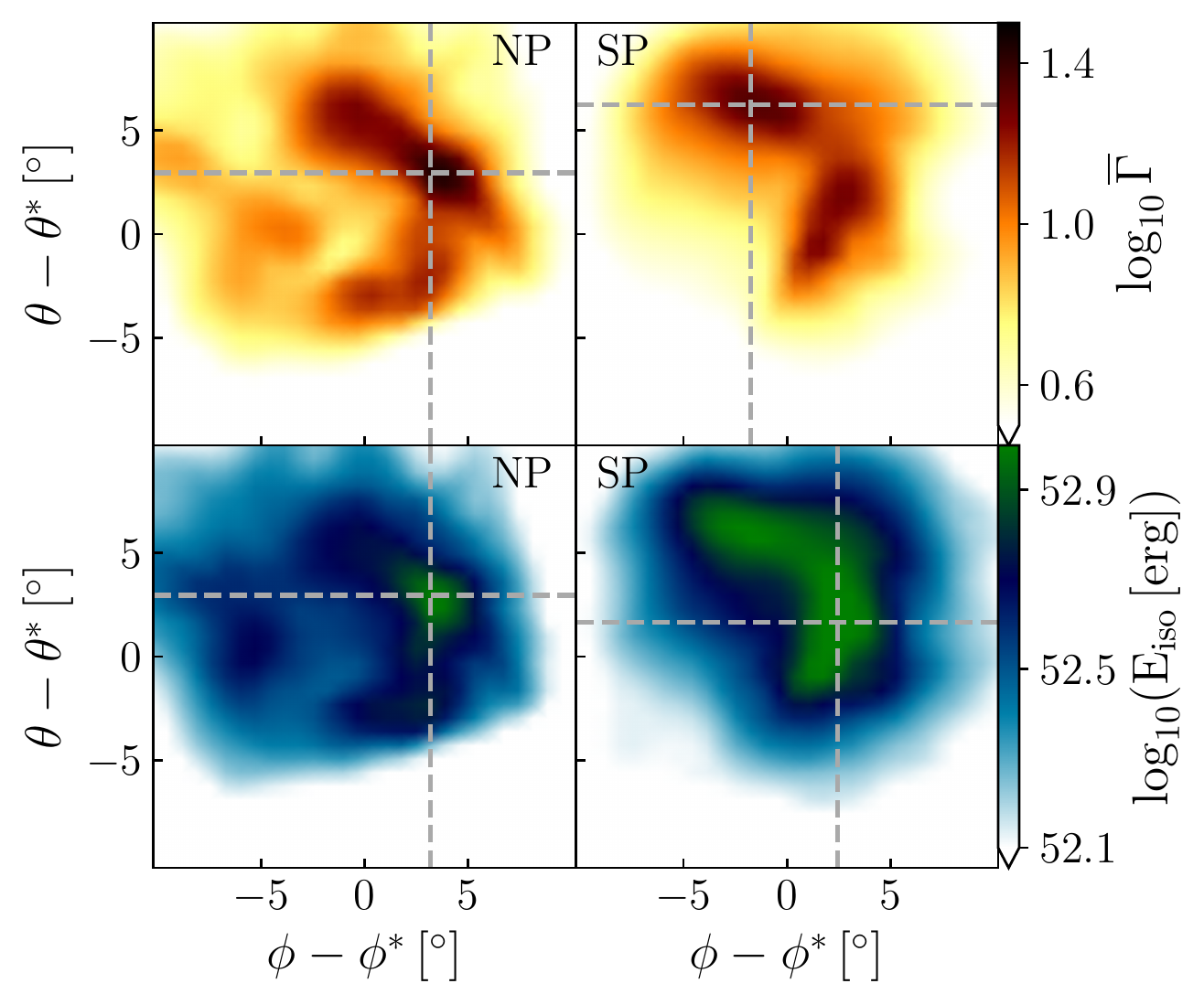}
    \caption{2D angular profiles of the jet head at $\mathrm{\simeq\!2385\,ms}$ after merger. In the upper panels, we show the distribution of radial-averaged Lorentz factor, from $\mathrm{\simeq\!3\!\times\!10^5\,km}$ to $\mathrm{\simeq\!6\!\times\!10^5\,km}$, weighted over the total energy density. In the lower panels, we show the corresponding isotropic equivalent energy distribution. On the left, values refer to the north pole of the jet, while on the right to the south pole (each distribution is computed relative to the $y\!>\!0$ axis, i.e.~$\mathrm{\mathnormal{\theta}^*\!=\!\mathnormal{\phi}^*\!=\!\pi/2}$). }
    \label{fig10}
\end{figure}

Our results show a first example of how a realistic magnetized BNS merger environment shapes the angular structure and energetics of an incipient sGRB jet (also magnetized).
Given the large extension and mass of its surroundings ($\mathrm{\gtrsim\!10^4\,km}$ and $\mathrm{\simeq\!0.22\,\mathnormal{M}_{\odot}}$, respectively), the incipient jet emerged with a highly collimated structure, characterized by a compact ultra-relativistic `head' with half-opening angle of $\mathrm{\simeq\!5^{\circ}}$, and maximum Lorentz factor and isotropic equivalent energy of $\mathrm{\simeq\!40}$ and $\mathrm{\simeq\!10^{53}\,erg}$, respectively (Figures\,\ref{fig9}-\ref{fig10}). 
The high degree of jet collimation and prolonged breakout time ($\mathrm{\simeq\!350\,ms}$ after launch) are a consequence of the extended launching time ($\mathrm{\simeq\!385\,ms}$ after merger) and the adopted injection configuration (Table\,\ref{inj_param}), as expected from other state-of-the-art jet propagation studies \citep[e.g.,][]{Geng2019,Hamidani2020,Murguia2021,Urrutia2023}. 
However, compared to the existing literature, we also took into account the effects of realistic anisotropies of the surrounding environment as well as a realistic and consistent magnetic field configuration, which significantly altered the overall jet propagation process.
In particular, inhomogeneities caused the jet to `snake' along paths of lower density, facilitating its emergence and generating significant 3D asymmetries in the large-scale angular structure (c.f.~Figure\,\ref{fig8}; see also \citealt{Lazzati2021}).
The presence of magnetic fields, on the other hand, made the jet-environment interface more stable and collimation more effective during propagation (notice the smooth jet edges at breakout in Figure\,\ref{fig8}), preventing the formation of significant turbulence as observed in similarly realistic but non-magnetized systems (P21). 

After the first 3D-RHD jet simulations employing realistic post-merger environments, reported in P21, the present work marks another fundamental quality step towards a more realistic and comprehensive description of sGRB jets in the context of BNS mergers, where magnetic fields represent a key ingredient.  
The approach demonstrated here provides the groundwork for exploring, in future investigations, the impact of different jet launching parameters (e.g., luminosity, magnetization, or jet launching time with respect to merger) on the overall propagation process that shapes the final observable jet signatures. 
Further extensions should eventually consider also a representative variety of reference BNS merger simulations. Finally, looking ahead, another major step forward in the consistency of our end-to-end description will be achieved when considering incipient jets directly produced in merger simulations.

\section*{Acknowledgements}

We thank the anonymous referee for constructive comments and suggestions. We also thank Luca Del Zanna for useful discussions.
This work is supported by the PRIN-INAF 2019 Grant ``Short gamma-ray burst jets from binary neutron star mergers'' (PI Ciolfi).
Simulations were performed on Galileo100 at CINECA (Italy) and on Discoverer at Sofia Tech Park (Bulgaria). In particular, we acknowledge CINECA for the availability of high performance computing resources and support through awards under the ISCRA initiative and the MoU INAF--CINECA (Grants IsC97\_MagJet, IsC97\_JetOut, INA21\_C8B57) and through a CINECA--INFN agreement (providing the allocation INF22\_teongrav).
Furthermore, we acknowledge the awarded EuroHPC Benchmark and Regular Access allocations on Discoverer EHPC-BEN-2022B02-038 and EHPC-REG-2022R03-218.

\section*{Data Availability}

The data underlying this article will be shared on reasonable request to the corresponding authors.



\bibliographystyle{mnras}
\bibliography{refs} 

\begin{thebibliography}{}
\makeatletter
\relax
\def\mn@urlcharsother{\let\do\@makeother \do\$\do\&\do\#\do\^\do\_\do\%\do\~}
\def\mn@doi{\begingroup\mn@urlcharsother \@ifnextchar [ {\mn@doi@}
  {\mn@doi@[]}}
\def\mn@doi@[#1]#2{\def\@tempa{#1}\ifx\@tempa\@empty \href
  {http://dx.doi.org/#2} {doi:#2}\else \href {http://dx.doi.org/#2} {#1}\fi
  \endgroup}
\def\mn@eprint#1#2{\mn@eprint@#1:#2::\@nil}
\def\mn@eprint@arXiv#1{\href {http://arxiv.org/abs/#1} {{\tt arXiv:#1}}}
\def\mn@eprint@dblp#1{\href {http://dblp.uni-trier.de/rec/bibtex/#1.xml}
  {dblp:#1}}
\def\mn@eprint@#1:#2:#3:#4\@nil{\def\@tempa {#1}\def\@tempb {#2}\def\@tempc
  {#3}\ifx \@tempc \@empty \let \@tempc \@tempb \let \@tempb \@tempa \fi \ifx
  \@tempb \@empty \def\@tempb {arXiv}\fi \@ifundefined
  {mn@eprint@\@tempb}{\@tempb:\@tempc}{\expandafter \expandafter \csname
  mn@eprint@\@tempb\endcsname \expandafter{\@tempc}}}

\bibitem[\protect\citeauthoryear{{Abbott} et~al.,}{{Abbott}
  et~al.}{2017a}]{LVC-BNS}
{Abbott} B.~P.,  et~al., 2017a, \mn@doi [Phys. Rev. Lett.]
  {10.1103/PhysRevLett.119.161101}, \href
  {http://adsabs.harvard.edu/abs/2017PhRvL.119p1101A} {119, 161101}

\bibitem[\protect\citeauthoryear{{Abbott} et~al.,}{{Abbott}
  et~al.}{2017b}]{LVC-Hubble}
{Abbott} B.~P.,  et~al., 2017b, \mn@doi [Nature] {10.1038/nature24471}, \href
  {http://adsabs.harvard.edu/abs/2017Natur.551...85A} {551, 85}

\bibitem[\protect\citeauthoryear{{Abbott} et~al.,}{{Abbott}
  et~al.}{2017c}]{LVC-MMA}
{Abbott} B.~P.,  et~al., 2017c, \mn@doi [Astrophys. J. Lett.]
  {10.3847/2041-8213/aa91c9}, \href
  {http://adsabs.harvard.edu/abs/2017ApJ...848L..12A} {848, L12}

\bibitem[\protect\citeauthoryear{{Abbott} et~al.,}{{Abbott}
  et~al.}{2017d}]{LVC-GRB}
{Abbott} B.~P.,  et~al., 2017d, \mn@doi [Astrophys. J. Lett.]
  {10.3847/2041-8213/aa920c}, \href
  {http://adsabs.harvard.edu/abs/2017ApJ...848L..13A} {848, L13}

\bibitem[\protect\citeauthoryear{{Abbott} et~al.,}{{Abbott}
  et~al.}{2019}]{LVC-170817properties}
{Abbott} B.~P.,  et~al., 2019, \mn@doi [Phys. Rev. X]
  {10.1103/PhysRevX.9.011001}, \href
  {https://ui.adsabs.harvard.edu/abs/2019PhRvX...9a1001A} {9, 011001}

\bibitem[\protect\citeauthoryear{Akmal, Pandharipande  \& Ravenhall}{Akmal
  et~al.}{1998}]{Akmal:1998:1804}
Akmal A.,  Pandharipande V.~R.,   Ravenhall D.~G.,  1998, \mn@doi [Phys. Rev.
  C] {10.1103/PhysRevC.58.1804}, 58, 1804

\bibitem[\protect\citeauthoryear{{Beniamini}, {Duran}, {Petropoulou}  \&
  {Giannios}}{{Beniamini} et~al.}{2020}]{Beniamini2020}
{Beniamini} P.,  {Duran} R.~B.,  {Petropoulou} M.,   {Giannios} D.,  2020,
  \mn@doi [Astrophys. J. Lett.] {10.3847/2041-8213/ab9223}, \href
  {https://ui.adsabs.harvard.edu/abs/2020ApJ...895L..33B} {895, L33}

\bibitem[\protect\citeauthoryear{{Bromberg}, {Nakar}, {Piran}  \&
  {Sari}}{{Bromberg} et~al.}{2011}]{Bromberg2011}
{Bromberg} O.,  {Nakar} E.,  {Piran} T.,   {Sari} R.,  2011, \mn@doi [\apj]
  {10.1088/0004-637X/740/2/100}, \href
  {https://ui.adsabs.harvard.edu/abs/2011ApJ...740..100B} {740, 100}

\bibitem[\protect\citeauthoryear{{Ciolfi}}{{Ciolfi}}{2020}]{Ciolfi2020a}
{Ciolfi} R.,  2020, \mn@doi [Mon. Not. R. Astron. Soc. Lett.]
  {10.1093/mnrasl/slaa062}, \href
  {https://ui.adsabs.harvard.edu/abs/2020MNRAS.495L..66C} {495, L66}

\bibitem[\protect\citeauthoryear{{Ciolfi} \& {Kalinani}}{{Ciolfi} \&
  {Kalinani}}{2020}]{CiolfiKalinani2020}
{Ciolfi} R.,  {Kalinani} J.~V.,  2020, \mn@doi [\apjl]
  {10.3847/2041-8213/abb240}, \href
  {https://ui.adsabs.harvard.edu/abs/2020ApJ...900L..35C} {900, L35}

\bibitem[\protect\citeauthoryear{{Ciolfi}, {Kastaun}, {Giacomazzo}, {Endrizzi},
  {Siegel}  \& {Perna}}{{Ciolfi} et~al.}{2017}]{Ciolfi2017}
{Ciolfi} R.,  {Kastaun} W.,  {Giacomazzo} B.,  {Endrizzi} A.,  {Siegel} D.~M.,
   {Perna} R.,  2017, \mn@doi [Phys. Rev. D] {10.1103/PhysRevD.95.063016},
  \href {http://adsabs.harvard.edu/abs/2017PhRvD..95f3016C} {95, 063016}

\bibitem[\protect\citeauthoryear{{Ciolfi}, {Kastaun}, {Kalinani}  \&
  {Giacomazzo}}{{Ciolfi} et~al.}{2019}]{Ciolfi2019}
{Ciolfi} R.,  {Kastaun} W.,  {Kalinani} J.~V.,   {Giacomazzo} B.,  2019,
  \mn@doi [Phys. Rev. D] {10.1103/PhysRevD.100.023005}, \href
  {https://ui.adsabs.harvard.edu/abs/2019PhRvD.100b3005C} {100, 023005}

\bibitem[\protect\citeauthoryear{{Dedner}, {Kemm}, {Kr{\"o}ner}, {Munz},
  {Schnitzer}  \& {Wesenberg}}{{Dedner} et~al.}{2002}]{Dedner2002}
{Dedner} A.,  {Kemm} F.,  {Kr{\"o}ner} D.,  {Munz} C.~D.,  {Schnitzer} T.,
  {Wesenberg} M.,  2002, \mn@doi [Journal of Computational Physics]
  {10.1006/jcph.2001.6961}, \href
  {https://ui.adsabs.harvard.edu/abs/2002JCoPh.175..645D} {175, 645}

\bibitem[\protect\citeauthoryear{{Eichler}, {Livio}, {Piran}  \&
  {Schramm}}{{Eichler} et~al.}{1989}]{Eichler1989}
{Eichler} D.,  {Livio} M.,  {Piran} T.,   {Schramm} D.~N.,  1989, \mn@doi
  [Nature] {10.1038/340126a0}, \href
  {http://adsabs.harvard.edu/abs/1989Natur.340..126E} {340, 126}

\bibitem[\protect\citeauthoryear{{Endrizzi}, {Ciolfi}, {Giacomazzo}, {Kastaun}
  \& {Kawamura}}{{Endrizzi} et~al.}{2016}]{Endrizzi2016}
{Endrizzi} A.,  {Ciolfi} R.,  {Giacomazzo} B.,  {Kastaun} W.,   {Kawamura} T.,
  2016, \mn@doi [Class. Quantum Grav.] {10.1088/0264-9381/33/16/164001}, \href
  {https://ui.adsabs.harvard.edu/abs/2016CQGra..33p4001E} {33, 164001}

\bibitem[\protect\citeauthoryear{{Garc{\'\i}a-Garc{\'\i}a},
  {L{\'o}pez-C{\'a}mara}  \& {Lazzati}}{{Garc{\'\i}a-Garc{\'\i}a}
  et~al.}{2023}]{Garcia-Garcia2023}
{Garc{\'\i}a-Garc{\'\i}a} L.,  {L{\'o}pez-C{\'a}mara} D.,   {Lazzati} D.,
  2023, \mn@doi [\mnras] {10.1093/mnras/stad023}, \href
  {https://ui.adsabs.harvard.edu/abs/2023MNRAS.519.4454G} {519, 4454}

\bibitem[\protect\citeauthoryear{{Geng}, {Zhang}, {K{\"o}lligan}, {Kuiper}  \&
  {Huang}}{{Geng} et~al.}{2019}]{Geng2019}
{Geng} J.-J.,  {Zhang} B.,  {K{\"o}lligan} A.,  {Kuiper} R.,   {Huang} Y.-F.,
  2019, \mn@doi [Astrophys. J. Lett.] {10.3847/2041-8213/ab224b}, \href
  {https://ui.adsabs.harvard.edu/abs/2019ApJ...877L..40G} {877, L40}

\bibitem[\protect\citeauthoryear{{Ghirlanda} et~al.,}{{Ghirlanda}
  et~al.}{2019}]{Ghirlanda2019}
{Ghirlanda} G.,  et~al., 2019, \mn@doi [Science] {10.1126/science.aau8815},
  \href {https://ui.adsabs.harvard.edu/abs/2019Sci...363..968G} {363, 968}

\bibitem[\protect\citeauthoryear{{Gill}, {Nathanail}  \& {Rezzolla}}{{Gill}
  et~al.}{2019}]{Gill2019}
{Gill} R.,  {Nathanail} A.,   {Rezzolla} L.,  2019, \mn@doi [Astrophys. J.]
  {10.3847/1538-4357/ab16da}, \href
  {https://ui.adsabs.harvard.edu/abs/2019ApJ...876..139G} {876, 139}

\bibitem[\protect\citeauthoryear{{Gottlieb} \& {Nakar}}{{Gottlieb} \&
  {Nakar}}{2022}]{Gottlieb2022b}
{Gottlieb} O.,  {Nakar} E.,  2022, \mn@doi [\mnras] {10.1093/mnras/stac2699},
  \href {https://ui.adsabs.harvard.edu/abs/2022MNRAS.517.1640G} {517, 1640}

\bibitem[\protect\citeauthoryear{{Gottlieb}, {Bromberg}, {Singh}  \&
  {Nakar}}{{Gottlieb} et~al.}{2020}]{Gottlieb2020}
{Gottlieb} O.,  {Bromberg} O.,  {Singh} C.~B.,   {Nakar} E.,  2020, \mn@doi
  [\mnras] {10.1093/mnras/staa2567}, \href
  {https://ui.adsabs.harvard.edu/abs/2020MNRAS.498.3320G} {498, 3320}

\bibitem[\protect\citeauthoryear{{Gottlieb}, {Nakar}  \& {Bromberg}}{{Gottlieb}
  et~al.}{2021}]{Gottlieb2021}
{Gottlieb} O.,  {Nakar} E.,   {Bromberg} O.,  2021, \mn@doi [Mon. Not. R.
  Astron.] {10.1093/mnras/staa3501}, \href
  {https://ui.adsabs.harvard.edu/abs/2021MNRAS.500.3511G} {500, 3511}

\bibitem[\protect\citeauthoryear{{Gottlieb}, {Moseley}, {Ramirez-Aguilar},
  {Murguia-Berthier}, {Liska}  \& {Tchekhovskoy}}{{Gottlieb}
  et~al.}{2022}]{Gottlieb2022a}
{Gottlieb} O.,  {Moseley} S.,  {Ramirez-Aguilar} T.,  {Murguia-Berthier} A.,
  {Liska} M.,   {Tchekhovskoy} A.,  2022, \mn@doi [\apjl]
  {10.3847/2041-8213/ac7728}, \href
  {https://ui.adsabs.harvard.edu/abs/2022ApJ...933L...2G} {933, L2}

\bibitem[\protect\citeauthoryear{{Hamidani}, {Kiuchi}  \& {Ioka}}{{Hamidani}
  et~al.}{2020}]{Hamidani2020}
{Hamidani} H.,  {Kiuchi} K.,   {Ioka} K.,  2020, \mn@doi [Mon. Not. R. Astron.
  Soc.] {10.1093/mnras/stz3231}, \href
  {https://ui.adsabs.harvard.edu/abs/2020MNRAS.491.3192H} {491, 3192}

\bibitem[\protect\citeauthoryear{{Harrison}, {Gottlieb}  \& {Nakar}}{{Harrison}
  et~al.}{2018}]{Harrison2018}
{Harrison} R.,  {Gottlieb} O.,   {Nakar} E.,  2018, \mn@doi [\mnras]
  {10.1093/mnras/sty760}, \href
  {https://ui.adsabs.harvard.edu/abs/2018MNRAS.477.2128H} {477, 2128}

\bibitem[\protect\citeauthoryear{{Kathirgamaraju}, {Tchekhovskoy}, {Giannios}
  \& {Barniol Duran}}{{Kathirgamaraju} et~al.}{2019}]{Kathirgamaraju2019}
{Kathirgamaraju} A.,  {Tchekhovskoy} A.,  {Giannios} D.,   {Barniol Duran} R.,
  2019, \mn@doi [Mon. Not. R. Astron. Soc.] {10.1093/mnrasl/slz012}, \href
  {https://ui.adsabs.harvard.edu/abs/2019MNRAS.484L..98K} {484, L98}

\bibitem[\protect\citeauthoryear{{Lamb}, {Nativi}, {Rosswog}, {Kann}, {Levan},
  {Lundman}  \& {Tanvir}}{{Lamb} et~al.}{2022}]{Lamb2022}
{Lamb} G.~P.,  {Nativi} L.,  {Rosswog} S.,  {Kann} D.~A.,  {Levan} A.,
  {Lundman} C.,   {Tanvir} N.,  2022, \mn@doi [Universe]
  {10.3390/universe8120612}, \href
  {https://ui.adsabs.harvard.edu/abs/2022Univ....8..612L} {8, 612}

\bibitem[\protect\citeauthoryear{{Lazzati}, {Perna}, {Morsony}, {Lopez-Camara},
  {Cantiello}, {Ciolfi}, {Giacomazzo}  \& {Workman}}{{Lazzati}
  et~al.}{2018}]{Lazzati2018}
{Lazzati} D.,  {Perna} R.,  {Morsony} B.~J.,  {Lopez-Camara} D.,  {Cantiello}
  M.,  {Ciolfi} R.,  {Giacomazzo} B.,   {Workman} J.~C.,  2018, \mn@doi [Phys.
  Rev. Lett.] {10.1103/PhysRevLett.120.241103}, \href
  {http://adsabs.harvard.edu/abs/2018PhRvL.120x1103L} {120, 241103}

\bibitem[\protect\citeauthoryear{{Lazzati}, {Ciolfi}  \& {Perna}}{{Lazzati}
  et~al.}{2020}]{Lazzati2020}
{Lazzati} D.,  {Ciolfi} R.,   {Perna} R.,  2020, \mn@doi [Astrophys. J.]
  {10.3847/1538-4357/ab9a44}, \href
  {https://ui.adsabs.harvard.edu/abs/2020ApJ...898...59L} {898, 59}

\bibitem[\protect\citeauthoryear{{Lazzati}, {Perna}, {Ciolfi}, {Giacomazzo},
  {L{\'o}pez-C{\'a}mara}  \& {Morsony}}{{Lazzati} et~al.}{2021}]{Lazzati2021}
{Lazzati} D.,  {Perna} R.,  {Ciolfi} R.,  {Giacomazzo} B.,
  {L{\'o}pez-C{\'a}mara} D.,   {Morsony} B.,  2021, \mn@doi [\apjl]
  {10.3847/2041-8213/ac1794}, \href
  {https://ui.adsabs.harvard.edu/abs/2021ApJ...918L...6L} {918, L6}

\bibitem[\protect\citeauthoryear{{Levinson} \& {Nakar}}{{Levinson} \&
  {Nakar}}{2020}]{LevisonNakar2020}
{Levinson} A.,  {Nakar} E.,  2020, \mn@doi [\physrep]
  {10.1016/j.physrep.2020.04.003}, \href
  {https://ui.adsabs.harvard.edu/abs/2020PhR...866....1L} {866, 1}

\bibitem[\protect\citeauthoryear{{Margutti} \& {Chornock}}{{Margutti} \&
  {Chornock}}{2021}]{Margutti2021}
{Margutti} R.,  {Chornock} R.,  2021, \mn@doi [\araa]
  {10.1146/annurev-astro-112420-030742}, \href
  {https://ui.adsabs.harvard.edu/abs/2021ARA&A..59..155M} {59, 155}

\bibitem[\protect\citeauthoryear{{Mart{\'\i}}}{{Mart{\'\i}}}{2015}]{Marti2015}
{Mart{\'\i}} J.-M.,  2015, \mn@doi [\mnras] {10.1093/mnras/stv1524}, \href
  {https://ui.adsabs.harvard.edu/abs/2015MNRAS.452.3106M} {452, 3106}

\bibitem[\protect\citeauthoryear{{Metzger}, {Thompson}  \&
  {Quataert}}{{Metzger} et~al.}{2018}]{Metzger2018}
{Metzger} B.~D.,  {Thompson} T.~A.,   {Quataert} E.,  2018, \mn@doi [Astrophys.
  J. Lett.] {10.3847/1538-4357/aab095}, \href
  {https://ui.adsabs.harvard.edu/abs/2018ApJ...856..101M} {856, 101}

\bibitem[\protect\citeauthoryear{{Mignone} \& {McKinney}}{{Mignone} \&
  {McKinney}}{2007}]{Mignone2007}
{Mignone} A.,  {McKinney} J.~C.,  2007, \mn@doi [Mon. Not. R. Astron. Soc.]
  {10.1111/j.1365-2966.2007.11849.x}, \href
  {https://ui.adsabs.harvard.edu/abs/2007MNRAS.378.1118M} {378, 1118}

\bibitem[\protect\citeauthoryear{{Mignone} \& {Tzeferacos}}{{Mignone} \&
  {Tzeferacos}}{2010}]{MT2010}
{Mignone} A.,  {Tzeferacos} P.,  2010, \mn@doi [Journal of Computational
  Physics] {10.1016/j.jcp.2009.11.026}, \href
  {https://ui.adsabs.harvard.edu/abs/2010JCoPh.229.2117M} {229, 2117}

\bibitem[\protect\citeauthoryear{{Mignone}, {Bodo}, {Massaglia}, {Matsakos},
  {Tesileanu}, {Zanni}  \& {Ferrari}}{{Mignone}
  et~al.}{2007}]{Mignone2007-PLUTO1}
{Mignone} A.,  {Bodo} G.,  {Massaglia} S.,  {Matsakos} T.,  {Tesileanu} O.,
  {Zanni} C.,   {Ferrari} A.,  2007, \mn@doi [Astrophys. J. Sup. Series]
  {10.1086/513316}, \href
  {https://ui.adsabs.harvard.edu/abs/2007ApJS..170..228M} {170, 228}

\bibitem[\protect\citeauthoryear{{Mignone}, {Tzeferacos}  \& {Bodo}}{{Mignone}
  et~al.}{2010}]{MTB2010}
{Mignone} A.,  {Tzeferacos} P.,   {Bodo} G.,  2010, \mn@doi [Journal of
  Computational Physics] {10.1016/j.jcp.2010.04.013}, \href
  {https://ui.adsabs.harvard.edu/abs/2010JCoPh.229.5896M} {229, 5896}

\bibitem[\protect\citeauthoryear{{Mignone}, {Zanni}, {Tzeferacos}, {van
  Straalen}, {Colella}  \& {Bodo}}{{Mignone} et~al.}{2012}]{Mignone2012-PLUTO2}
{Mignone} A.,  {Zanni} C.,  {Tzeferacos} P.,  {van Straalen} B.,  {Colella} P.,
    {Bodo} G.,  2012, \mn@doi [Astrophys. J. Sup. Series]
  {10.1088/0067-0049/198/1/7}, \href
  {https://ui.adsabs.harvard.edu/abs/2012ApJS..198....7M} {198, 7}

\bibitem[\protect\citeauthoryear{{Mooley} et~al.,}{{Mooley}
  et~al.}{2018}]{Mooley2018b}
{Mooley} K.~P.,  et~al., 2018, \mn@doi [Nature] {10.1038/s41586-018-0486-3},
  \href {http://adsabs.harvard.edu/abs/2018Natur.561..355M} {561, 355}

\bibitem[\protect\citeauthoryear{{Mooley}, {Anderson}  \& {Lu}}{{Mooley}
  et~al.}{2022}]{Mooley2022}
{Mooley} K.~P.,  {Anderson} J.,   {Lu} W.,  2022, \mn@doi [\nat]
  {10.1038/s41586-022-05145-7}, \href
  {https://ui.adsabs.harvard.edu/abs/2022Natur.610..273M} {610, 273}

\bibitem[\protect\citeauthoryear{{Murguia-Berthier}, {Ramirez-Ruiz}, {De
  Colle}, {Janiuk}, {Rosswog}  \& {Lee}}{{Murguia-Berthier}
  et~al.}{2021}]{Murguia2021}
{Murguia-Berthier} A.,  {Ramirez-Ruiz} E.,  {De Colle} F.,  {Janiuk} A.,
  {Rosswog} S.,   {Lee} W.~H.,  2021, \mn@doi [Astrophys. J.]
  {10.3847/1538-4357/abd08e}, \href
  {https://ui.adsabs.harvard.edu/abs/2021ApJ...908..152M} {908, 152}

\bibitem[\protect\citeauthoryear{{Nakar}}{{Nakar}}{2020}]{Nakar2020}
{Nakar} E.,  2020, \mn@doi [Phys. Rep.] {10.1016/j.physrep.2020.08.008}, \href
  {https://ui.adsabs.harvard.edu/abs/2020PhR...886....1N} {886, 1}

\bibitem[\protect\citeauthoryear{{Nathanail}, {Gill}, {Porth}, {Fromm}  \&
  {Rezzolla}}{{Nathanail} et~al.}{2020}]{Nathanail2020}
{Nathanail} A.,  {Gill} R.,  {Porth} O.,  {Fromm} C.~M.,   {Rezzolla} L.,
  2020, \mn@doi [Mon. Not. R. Astron. Soc.] {10.1093/mnras/staa1454}, \href
  {https://ui.adsabs.harvard.edu/abs/2020MNRAS.495.3780N} {495, 3780}

\bibitem[\protect\citeauthoryear{{Nathanail}, {Gill}, {Porth}, {Fromm}  \&
  {Rezzolla}}{{Nathanail} et~al.}{2021}]{Nathanail2021}
{Nathanail} A.,  {Gill} R.,  {Porth} O.,  {Fromm} C.~M.,   {Rezzolla} L.,
  2021, \mn@doi [Mon. Not. R. Astron. Soc.] {10.1093/mnras/stab115}, \href
  {https://ui.adsabs.harvard.edu/abs/2021MNRAS.502.1843N} {502, 1843}

\bibitem[\protect\citeauthoryear{{Nativi}, {Bulla}, {Rosswog}, {Lundman},
  {Kowal}, {Gizzi}, {Lamb}  \& {Perego}}{{Nativi} et~al.}{2021}]{Nativi2021}
{Nativi} L.,  {Bulla} M.,  {Rosswog} S.,  {Lundman} C.,  {Kowal} G.,  {Gizzi}
  D.,  {Lamb} G.~P.,   {Perego} A.,  2021, \mn@doi [Mon. Not. R. Astron. Soc.]
  {10.1093/mnras/staa3337}, \href
  {https://ui.adsabs.harvard.edu/abs/2021MNRAS.500.1772N} {500, 1772}

\bibitem[\protect\citeauthoryear{{Nativi}, {Lamb}, {Rosswog}, {Lundman}  \&
  {Kowal}}{{Nativi} et~al.}{2022}]{Nativi2022}
{Nativi} L.,  {Lamb} G.~P.,  {Rosswog} S.,  {Lundman} C.,   {Kowal} G.,  2022,
  \mn@doi [\mnras] {10.1093/mnras/stab2982}, \href
  {https://ui.adsabs.harvard.edu/abs/2022MNRAS.509..903N} {509, 903}

\bibitem[\protect\citeauthoryear{{Paczynski}}{{Paczynski}}{1986}]{Paczynski1986}
{Paczynski} B.,  1986, \mn@doi [Astrophys. J. Lett.] {10.1086/184740}, \href
  {http://adsabs.harvard.edu/abs/1986ApJ...308L..43P} {308, L43}

\bibitem[\protect\citeauthoryear{{Pavan}, {Ciolfi}, {Kalinani}  \&
  {Mignone}}{{Pavan} et~al.}{2021}]{Pavan2021}
{Pavan} A.,  {Ciolfi} R.,  {Kalinani} J.~V.,   {Mignone} A.,  2021, \mn@doi
  [\mnras] {10.1093/mnras/stab1810}, \href
  {https://ui.adsabs.harvard.edu/abs/2021MNRAS.506.3483P} {506, 3483}

\bibitem[\protect\citeauthoryear{{Ruiz}, {Lang}, {Paschalidis}  \&
  {Shapiro}}{{Ruiz} et~al.}{2016}]{Ruiz2016}
{Ruiz} M.,  {Lang} R.~N.,  {Paschalidis} V.,   {Shapiro} S.~L.,  2016, \mn@doi
  [Astrophys. J. Lett.] {10.3847/2041-8205/824/1/L6}, \href
  {http://adsabs.harvard.edu/abs/2016ApJ...824L...6R} {824, L6}

\bibitem[\protect\citeauthoryear{{Salafia} \& {Ghirlanda}}{{Salafia} \&
  {Ghirlanda}}{2022}]{Salafia2022}
{Salafia} O.~S.,  {Ghirlanda} G.,  2022, \mn@doi [Galaxies]
  {10.3390/galaxies10050093}, \href
  {https://ui.adsabs.harvard.edu/abs/2022Galax..10...93S} {10, 93}

\bibitem[\protect\citeauthoryear{{Sun}, {Ruiz}, {Shapiro}  \& {Tsokaros}}{{Sun}
  et~al.}{2022}]{Sun2022}
{Sun} L.,  {Ruiz} M.,  {Shapiro} S.~L.,   {Tsokaros} A.,  2022, \mn@doi [\prd]
  {10.1103/PhysRevD.105.104028}, \href
  {https://ui.adsabs.harvard.edu/abs/2022PhRvD.105j4028S} {105, 104028}

\bibitem[\protect\citeauthoryear{{Urrutia}, {De Colle}, {Murguia-Berthier}  \&
  {Ramirez-Ruiz}}{{Urrutia} et~al.}{2021}]{Urrutia2021}
{Urrutia} G.,  {De Colle} F.,  {Murguia-Berthier} A.,   {Ramirez-Ruiz} E.,
  2021, \mn@doi [Mon. Not. R. Astron. Soc.] {10.1093/mnras/stab723}, \href
  {https://ui.adsabs.harvard.edu/abs/2021MNRAS.tmp..723U} {}

\bibitem[\protect\citeauthoryear{{Urrutia}, {De Colle}  \&
  {L{\'o}pez-C{\'a}mara}}{{Urrutia} et~al.}{2023}]{Urrutia2023}
{Urrutia} G.,  {De Colle} F.,   {L{\'o}pez-C{\'a}mara} D.,  2023, \mn@doi
  [\mnras] {10.1093/mnras/stac3401}, \href
  {https://ui.adsabs.harvard.edu/abs/2023MNRAS.518.5145U} {518, 5145}

\bibitem[\protect\citeauthoryear{{Xie}, {Zrake}  \& {MacFadyen}}{{Xie}
  et~al.}{2018}]{Xie2018}
{Xie} X.,  {Zrake} J.,   {MacFadyen} A.,  2018, \mn@doi [Astrophys. J.]
  {10.3847/1538-4357/aacf9c}, \href
  {https://ui.adsabs.harvard.edu/abs/2018ApJ...863...58X} {863, 58}

\bibitem[\protect\citeauthoryear{{Zhang}}{{Zhang}}{2019}]{Zhang2019}
{Zhang} B.,  2019, \mn@doi [Front. Phys.] {10.1007/s11467-019-0913-4}, \href
  {https://ui.adsabs.harvard.edu/abs/2019FrPhy..1464402Z} {14, 64402}

\makeatother
\end{thebibliography}


\appendix

\section{Divergence cleaning setup}
\label{cleaning}

To ensure the divergence-free condition for the magnetic field in our simulations, we employ the numerical technique by \cite{Dedner2002} already implemented in \textsc{PLUTO} \citep{MT2010,MTB2010}. This involves a modified system of conservation laws in which a generalized multiplier is coupled to the induction equation
\begin{equation}\label{induction}
    \partial_t\vec{B}-\vec{\nabla}\times\left(\vec{v}\times\vec{B}\right)=0\, ,
\end{equation}
causing divergence errors to be transported to the domain boundaries at maximal admissible speed, and damped at the same time. In particular, the new system of conservation laws is still hyperbolic, and the density, momentum, magnetic induction, and total energy density are still conserved.

\begin{figure}
	\includegraphics[width=\columnwidth,keepaspectratio]{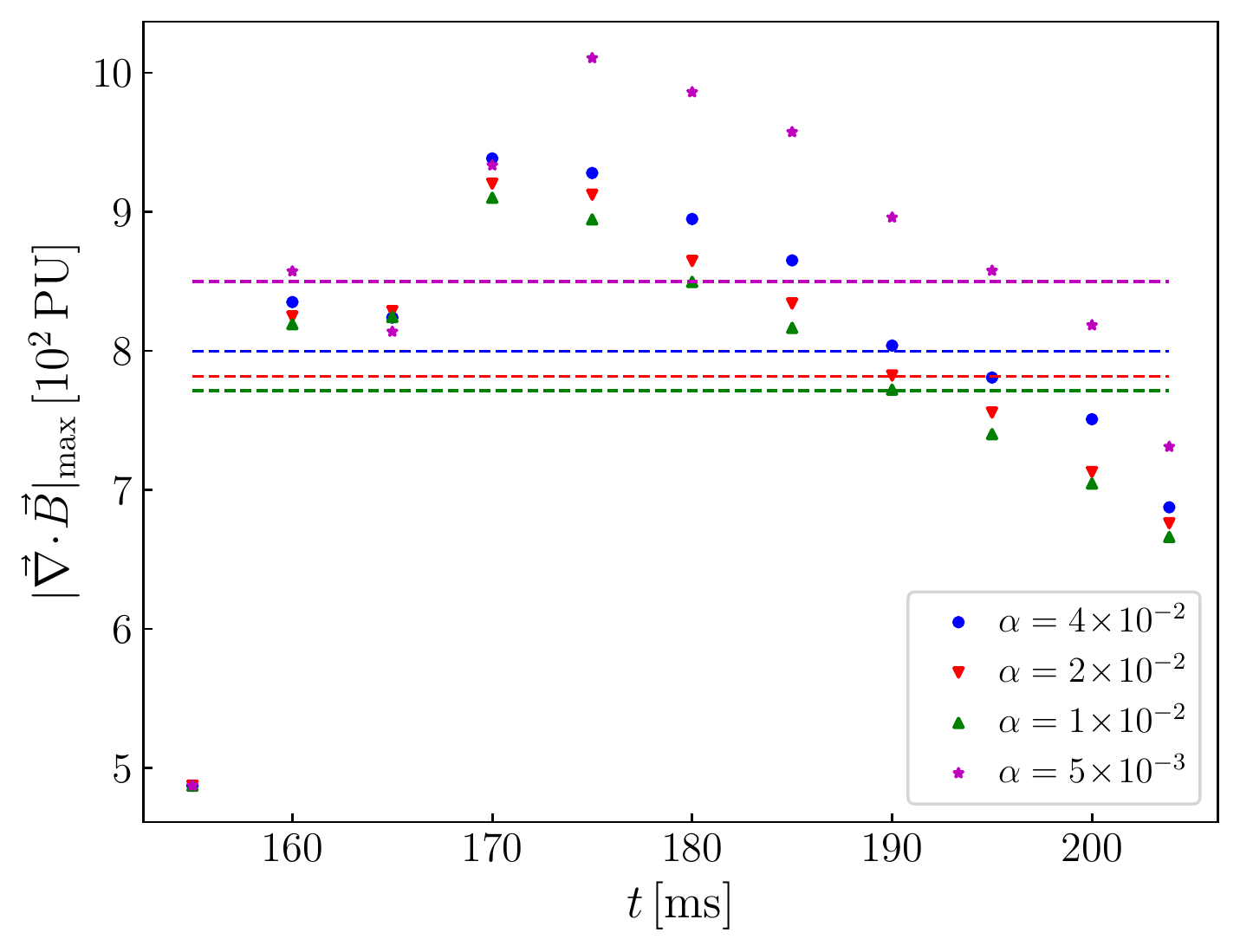}
    \caption{Time behaviour of the maximum divergence of the magnetic field (in \textsc{pluto}'s units, i.e., $\mathrm{\simeq\!7.2\!\times\!10^4\,G/cm}$) for different values of $\alpha$. Each trend refers to the same initial system evolved for 50\,ms after the data import (i.e., from $\simeq\!155$\,ms to $\simeq\!205$\,ms after merger). The average value of $|\vec{\nabla}\!\cdot\!\vec{B}|_{\mathrm{max}}$ for each test performed is shown through dashed horizontal lines. See text for additional details.}
    \label{divB}
\end{figure}

Within such a new formulation, the induction equation and the solenoidal constraint are replaced, respectively, by
\begin{equation}\label{GLM1}
    \partial_t\vec{B}-\vec{\nabla}\times\left(\vec{v}\times\vec{B}\right)+\vec{\nabla}\psi=0
\end{equation}
and
\begin{equation}\label{GLM2}
    \partial_t\psi+c^2_h\vec{\nabla}\cdot\vec{B}=-\dfrac{c^2_h}{c^2_p}\psi\, ,
\end{equation}
where $c_h\!=\!\mathrm{CFL}\!\times\!\Delta l_{\mathrm{min}}/\Delta t$ is the maximum speed compatible with the step size $\Delta t$, $c_p\!=\!\sqrt{\Delta l_{\mathrm{min}}c_h/\alpha}$, and $\Delta l_{\mathrm{min}}$ and CFL are the minimum cell length and the Courant-Friedrichs-Lewy number, respectively. The damping rate of magnetic monopoles is controlled by the user-defined constant $\alpha$, which we set in our simulation by means of appropriate numerical tests. In particular, we simulate the first 50\,ms of evolution of our system (using the procedure described in Section\,\ref{extr}), and analyze the trend of the magnetic field divergence for different values of $\alpha$. The results of this analysis are illustrated in Figure\,\ref{divB}. We find that $\alpha\!=\!0.01$ is the most appropriate value for the system at hand, so we adopt it in all our simulations.

\section{Jet injection recipe}
\label{jet_calc}

In this Appendix, we provide more details about our jet prescription, with the aim of better clarifying to the reader the several steps discussed in Section\,\ref{jet}. In particular, in \ref{transv_eq} we describe the transverse equilibrium condition, and the calculations performed in order to derive the jet co-moving density and pressure. Then, in \ref{time_pres} we discuss the various assumptions made to set the time dependence of the jet. Finally, in \ref{vec_tilt} we report the calculations done for the $90^{\circ}$ coordinate tilting. 

\subsection{Transverse equilibrium}\label{transv_eq}

Within the framework of ideal RMHD, the conservation of energy-momentum reads
\begin{equation}\label{ener-mom}
    \partial_{\mu}T^{\nu\mu}+\Tilde{\Gamma}^{\nu}_{\sigma\mu}T^{\sigma\mu}+\Tilde{\Gamma}^{\mu}_{\sigma\mu}T^{\nu\sigma} = 0 \, ,
\end{equation}
where, assuming $c\!=\!1$,
\begin{equation}
    T^{\alpha\beta} = g^{\alpha\beta}\left(P+\dfrac{b^2}{2}\right)+\rho h^*u^{\alpha}u^{\beta}-b^{\alpha}b^{\beta} 
\end{equation}
is the energy-momentum tensor, and $\Tilde{\Gamma}^{\alpha\beta}_{\gamma}$ are the Christoffel symbols related to the metric $g_{\alpha\beta}$ (see Section\,\ref{jet} for the definition of the remaining physical variables).

In 3D-spherical coordinates, the above leads to the following transverse balance equation between total pressure gradient, centrifugal force, and magnetic tension:
\begin{align}
\begin{split}
    &\partial_r\left(\dfrac{\Tilde{T}^{\theta r}}{r}\right)+\partial_{\theta}\left(\dfrac{\Tilde{T}^{\theta\theta}}{r^2}\right)+\partial_{\phi}\left(\dfrac{\Tilde{T}^{\theta\phi}}{r^2\st}\right)\\
    &+\dfrac{1}{r^2}\left(2+\dfrac{1}{\st}\right)\Tilde{T}^{r\theta}-\st\ct\dfrac{1}{r^2\st^2}\Tilde{T}^{\phi\phi}+\dfrac{\ct}{r^2\st^2}\Tilde{T}^{\theta\phi}=0
\end{split}
\end{align}
where $\st\!=\!\sin{\theta}$, $\ct\!=\!\cos{\theta}$, and $\Tilde{T}$ is normalized according to 
\begin{equation}
    \Tilde{T}^{\alpha\beta}=T^{\alpha\beta}\sqrt{g_{\alpha\alpha}g_{\beta\beta}} \, .
\end{equation}
Since $v^{\theta}\!=\!B^{\theta}\!=\!0$ in our jet prescription, we can rewrite the balance equation as
\begin{equation}
    \dfrac{d}{d\theta}\left(P+\dfrac{b^2}{2}\right)-\dfrac{\ct}{\st}\left[\rho h^*\Gamma^2(v^{\phi})^2+P+\dfrac{b^2}{2}-(b^{\phi})^2\right]=0 \,,
\end{equation}
in which 
\begin{equation}
    b^{\phi} = \dfrac{1}{\sqrt{4\pi}}\left[\dfrac{B^{\phi}}{\Gamma}+\Gamma v^{\phi}(\Vec{v}\cdot\Vec{B})\right]\,.
\end{equation}

Then, by taking $\Gamma_{\mathrm{ad}}\!=\!4/3$, we get the following differential equation for the co-moving density of the jet:
\begin{equation}
    \dfrac{d}{d\theta}\rho-\dfrac{4h^*\Gamma^2(v^{\phi})^2+h^*-1}{\st(h^*-1)}\rho=-\dfrac{4(b^{\phi})^2-b^2+\st\dfrac{db^2}{d\theta}}{\st(h^*-1)} \, ;
\end{equation}
which can be solved using the general formula 
\begin{align}\label{rho_sol}
\begin{split}
    \rho(\theta)&=\rho(\theta_{\mathrm{j}})\exp\left[-\int_{\theta_{\mathrm{j}}}^{\theta}f(\theta')d\theta'\right]+\exp\left[-\int_{\theta_{\mathrm{j}}}^{\theta}f(\theta')d\theta'\right]\\
    &\times\int_{\theta_{\mathrm{j}}}^{\theta}\left\{\exp\left[\int_{\theta_{\mathrm{j}}}^{\zeta}f(\theta')d\theta'\right]g(\zeta)d\zeta\right\} \, ,
\end{split}
\end{align}
where
\begin{align}
\begin{split}
    \int_{\theta_{\mathrm{j}}}^{\theta}f(\theta')d\theta'&=\int_{\theta_{\mathrm{j}}}^{\theta}\left[-\dfrac{4h^*\Gamma^2(v^{\phi})^2+h^*-1}{\st(h^*-1)}\right]d\theta'\\
    &\simeq-\dfrac{4h^*\Gamma^2\overline{\Omega}^2\mathrm{r^2_{exc}}}{h^*-1}\dfrac{\theta^2-\theta_{\mathrm{j}}^2}{2}-\ln{\dfrac{\theta}{\theta_{\mathrm{j}}}} \, ,
\end{split}
\end{align}
and
\begin{equation}
    g(\theta)=-\dfrac{4(b^{\phi})^2-b^2+\st\dfrac{db^2}{d\zeta}}{\st(h^*-1)} \, .
\end{equation}
In particular, we compute numerically the last integral of Eq.\,\ref{rho_sol}, and fix the `initial' value $\rho(\theta_{\mathrm{j}})$ using the jet luminosity defined in Eq.~\ref{jet_lum}. Such a solution, also provides us with the angular profile for the jet pressure from Eq.~\ref{jet_prs}.

\subsection{Time prescription}\label{time_pres}

To model the time dependence of our jet, we start by considering the definition of jet luminosity,  
\begin{equation}\label{jet_lum2}
    L=4\pi r^2_\mathrm{exc}\int_0^{\theta_j}\left[\rho h^*\Gamma^2c^2-\left(P+\dfrac{b^2}{2}\right)-(b^0)^2\right]v^r\st d\theta \,,
\end{equation}
and attempt to vary it as $L(t)\!=\!L_{\mathrm{j}}e^{-t/\tau_{\mathrm{d}}}$, where $\tau_{\mathrm{d}}\!=\!0.3$\,s is the timescale of the accretion process assumed to power the jet (see Section\,\ref{collapse}). By denoting $t_\mathrm{in}$ as the time of jet injection, we set
\begin{align}
    &\rho(t)=\rho(t_\mathrm{in}) \, ,\\
    &v^{\phi}(t)=v^{\phi}(t_\mathrm{in}) \, ,
\end{align}
and 
\begin{align}
    &\{h^*\Gamma\}(t)=\{h^*\Gamma\}(t_\mathrm{in})\,e^{-t/2\tau_{\mathrm{d}}} \, ,\\
    &\{\Gamma v^r\}(t)=\{\Gamma v^r\}(t_\mathrm{in})\,e^{-t/2\tau_{\mathrm{d}}} \, ,
\end{align}
which yields $\{\rho h^*\Gamma^2v^r\}(t)\!=\!\{\rho h^*\Gamma^2v^r\}(t_\mathrm{in})\,e^{-t/\tau_{\mathrm{d}}}$, along with Eq.~\ref{vr_t}, \ref{Gm_t}, \ref{h*_t}. Moreover, given the definition of $h^*$ (Eq.\,\ref{h^*}), we take
\begin{align}
    &\{\rho h\Gamma^2v^r\}(t)=\{\rho h\Gamma^2v^r\}(t_\mathrm{in})\,e^{-t/\tau_{\mathrm{d}}} \, ,\\
    &\{b^2\Gamma^2v^r\}(t)\!=\!\{b^2\Gamma^2v^r\}(t_\mathrm{in})\,e^{-t/\tau_{\mathrm{d}}} \label{eqB1}\,, 
\end{align}
obtaining
\begin{equation}
    \left\{\dfrac{b^2}{2}v^r\right\}(t)=\left\{\dfrac{b^2}{2}v^r\right\}(t_\mathrm{in})\dfrac{\Gamma^2(t_\mathrm{in})}{\Gamma^2(t)}\,e^{-t/\tau_{\mathrm{d}}} \, ,
\end{equation}
and 
\begin{align}
\begin{split}
    \{P v^r\}(t)&=\left\{\dfrac{1}{4}\left[\rho(h^*-1)-b^2\right]v^r\right\}(t)\\
    &=\dfrac{1}{4}\left[\dfrac{\{\rho h^*\Gamma^2v^r-b^2\Gamma^2v^r\}(t_\mathrm{in})\,e^{-t/\tau_{\mathrm{d}}}}{\Gamma^2(t)}-\rho\right]\\
    &\simeq\{Pv^r\}(t_\mathrm{in})\dfrac{\Gamma^2(t_\mathrm{in})}{\Gamma^2(t)}\,e^{-t/\tau_{\mathrm{d}}}\, ,
\end{split}
\end{align}
which implies a temporal variation of the second addend in Eq.\,\ref{jet_lum2} as $\simeq\!\Gamma^2(t_\mathrm{in})e^{-t/\tau_{\mathrm{d}}}/\Gamma^2(t)$.

Finally, given the targeted time dependence of the jet, we set
\begin{equation}\label{eqB2}
    \{(b^0)^2v^r\}(t)=\{(b^0)^2v^r\}(t_\mathrm{in})e^{-t/\tau_{\mathrm{d}}}\, ,
\end{equation}
which, by combining Eq.~\ref{b2}, \ref{b0}, \ref{eqB1}, \ref{eqB2}, gives us
\begin{align}
\begin{split}
    B^2(t)&=4\pi\{\Gamma^2b^2-(b^0)^2\}(t)\\
    &=4\pi\{\Gamma^2b^2-(b^0)^2\}(t_\mathrm{in})\dfrac{v^r(t_\mathrm{in})}{v^r(t)}e^{-t/\tau_{\mathrm{d}}}\\
    &=B^2(t_\mathrm{in})\dfrac{v^r(t_\mathrm{in})}{v^r(t)}e^{-t/\tau_{\mathrm{d}}}\, ,
\end{split}
\end{align}
and thus the time profiles of the magnetic fields reported in Eqs.~\ref{Bphi_t}, \ref{Br_t}. 

\begin{figure}
	\includegraphics[width=\columnwidth,keepaspectratio]{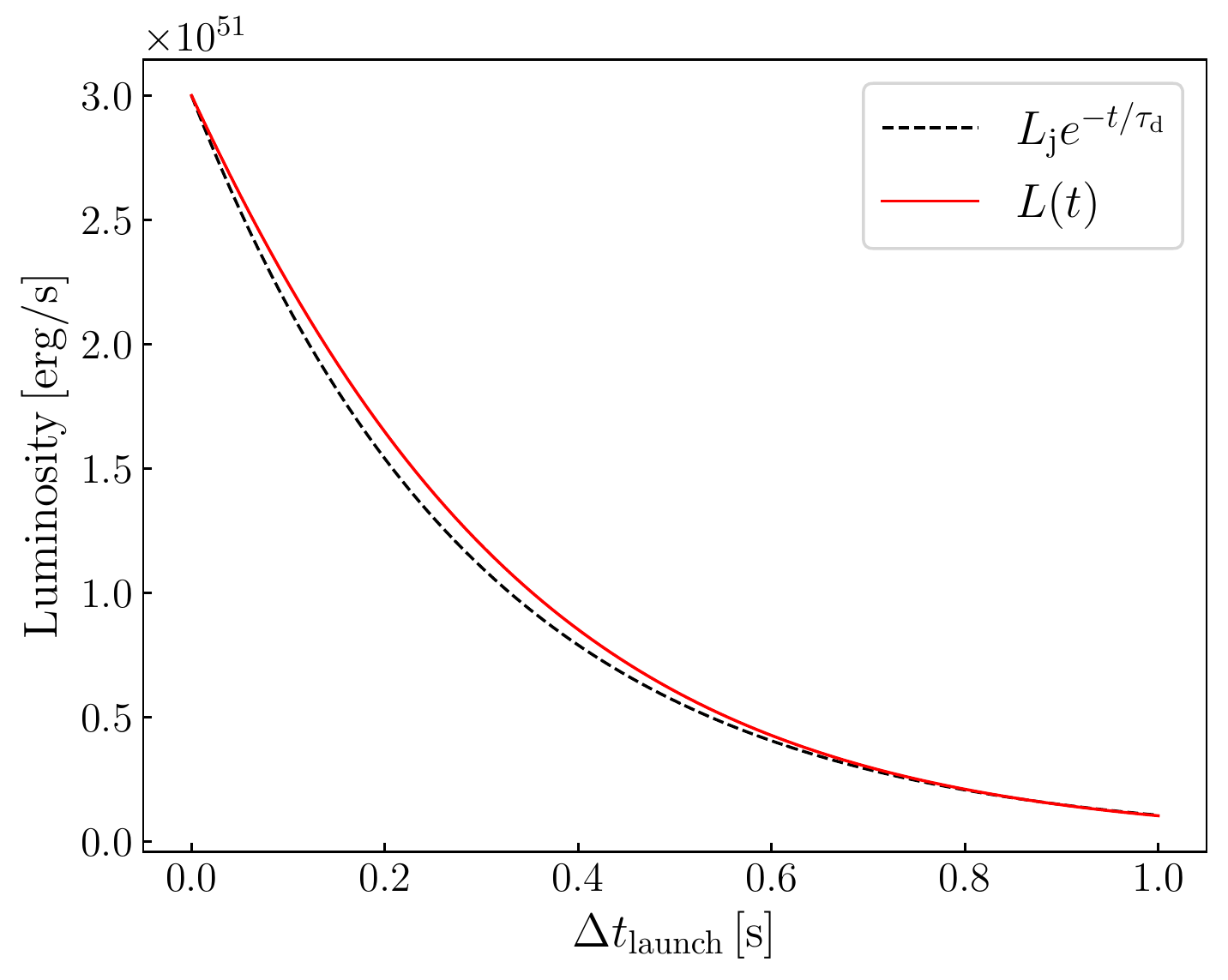}
    \caption{Targeted variation of jet luminosity over time (dashed-gray curve) compared with the actual trend yielded by our time prescription (red curve). The exponential function that best fits the actual trend is $L(t)\!=\!L_{\mathrm{j}}e^{-t/\tau'}$, where $\tau'\!\simeq\!0.307$\,s (very close to $\tau_{\mathrm{d}}\!=\!0.3$\,s).}.
    \label{jet_lum3}
\end{figure}

The above time prescription leads to a jet luminosity that varies over time as shown in Figure\,\ref{jet_lum3} (red curve). In particular, we find that the exponential function that best fits this trend is $L(t)\!=\!L_{\mathrm{j}}e^{-t/\tau'}$, where $\mathrm{\mathnormal{\tau'}\!\simeq\!0.307\,s\,(\approx\!\mathnormal{\tau}_{d}})$.

\subsection{Tilting of vector fields}\label{vec_tilt}

To arrange the jet with the main axis of symmetry parallel to the orbital $y$-axis of the system, it is necessary to tilt the jet fields given by our prescription, since these are obtained by assuming the jet axis parallel to the coordinate $z$-axis. 

To do this, we first convert the spherical basis $\Vec{e}_r,\Vec{e}_{\theta},\Vec{e}_{\phi}$ to Cartesian coordinates, i.e., 
\begin{align}
\begin{cases}
    \Vec{e}_r = \st\cp\,\Vec{e}_x+\st\sp\,\Vec{e}_y+\ct\,\Vec{e}_z \\
    \Vec{e}_{\theta} = \ct\cp\,\Vec{e}_x+\ct\sp\,\Vec{e}_y-\st\,\Vec{e}_z \\
    \Vec{e}_{\phi} = -\sp\,\Vec{e}_x+\cp\,\Vec{e}_y
\end{cases}\ ,
\end{align}
where $\sp\!=\!\sin{\phi}$ and $\cp\!=\!\cos{\phi}$. Thus, being $\Vec{B}\!=\!(B^r,0,B^{\phi})$ the magnetic field of the jet (see Eq.\,\ref{BphiBr}), we have
\begin{align}
\begin{split}
    \Vec{B} &= B^{\phi}(-\sp\,\Vec{e}_x+\cp\,\Vec{e}_y)+B^{r}(\st\cp\,\Vec{e}_x+\st\sp\,\Vec{e}_y+\ct\,\Vec{e}_z)\\
    &= (-B^{\phi}\sp+B^r\st\cp)\,\Vec{e}_x+(B^{\phi}\cp+B^r\st\sp)\,\Vec{e}_y\\
    &+(B^r\ct)\,\Vec{e}_z\ .
\end{split}
\end{align}
Moreover, by tilting the Cartesian basis as  
\begin{align}
\begin{cases}
    \Vec{e}_{x} \to \Vec{e}_{x'} \\
    \Vec{e}_{y} \to -\Vec{e}_{z'} \\
    \Vec{e}_{z} \to \Vec{e}_{y'}
\end{cases}\ ,
\end{align}
where the superscript $'$ denotes quantities in the tilted frame, we obtain a new expression of $\Vec{B}$ in the tilted frame, i.e., 
\begin{align}\label{Bxyz_tilt}
\begin{split}
     \Vec{B} &= (-B^{\phi}\sp+B^r\st\cp)\,\Vec{e}_{x'}-(B^{\phi}\cp+B^r\st\sp)\,\Vec{e}_{z'}\\
     &+(B^r\ct)\,\Vec{e}_{y'}\ .
\end{split}
\end{align}
After this, going back to spherical coordinates, i.e.,
\begin{equation}
\begin{cases}
    \Vec{e}_{x'} = \stp\cpp\,\Vec{e}_{r'}+\ctp\cpp\,\Vec{e}_{\theta'}-\spp\,\Vec{e}_{\phi'} \\
    \Vec{e}_{y'} = \stp\spp\,\Vec{e}_{r'}+\ctp\spp\,\Vec{e}_{\theta'}+\cpp\,\Vec{e}_{\phi'} \\
    \Vec{e}_{z'} = \ctp\,\Vec{e}_{r'}-\stp\,\Vec{e}_{\theta'}
\end{cases}\ ,
\end{equation}
we can rewrite Eq.~\ref{Bxyz_tilt} as 
\begin{align}
\begin{split}
    \Vec{B} =& [-B^{\phi}(\sp\stp\cpp+\cp\ctp)\\
    &+B^r(\st\cp\stp\cpp-\st\sp\ctp+\ct\stp\spp)]\,\Vec{e}_{r'}\\
    +& [-B^{\phi}(\sp\ctp\cpp-\cp\stp)\\
    &+B^r(\st\cp\ctp\cpp+\st\sp\stp+\ct\ctp\spp)]\,\Vec{e}_{\theta'}\\
    +& [B^{\phi}(\sp\spp)+B^r(-\st\cp\spp+\ct\cpp)]\,\Vec{e}_{\phi'}\ .
\end{split}
\end{align}
Finally, after some trigonometric calculations, we arrive at the expression of $\Vec{B}$ in spherical coordinates, within the tilted frame, which we employed in our simulations, i.e.,
\begin{equation}
    \Vec{B} = B^r\,\Vec{e}_{r'} + B^{\phi}\dfrac{\cos{\phi}}{\sin{\theta'}}\,\Vec{e}_{\theta'}+B^{\phi}\sin{\phi}\sin{\phi'}\Vec{e}_{\phi'}\,, 
\end{equation}
with 
\begin{equation}
    \phi = -\dfrac{|\cos{\theta'}|}{\cos{\theta'}}\arccos{\left(\dfrac{|\cos{\phi'}\sin{\theta'}|}{\cos{\phi'}\sin{\theta'}}\dfrac{1}{\sqrt{1+(\xi')^2}}\right)}\,,
\end{equation}
where $\xi'\!=\!(\cos{\phi'}\tan{\theta'})^{-1}$.
Notice that now
\begin{equation}
    B^{\phi}=\dfrac{2B^{\phi}_{\mathrm{j,m}}(\alpha/\theta_{\mathrm{j,m}})}{1+(\alpha/\theta_{\mathrm{j,m}})^2}\, ,
\end{equation}
where $\alpha$ is the angle with respect to the $y$-axis.
\section{Impact of EOS choice}
Setting an appropriate EOS for our post-merger system, including the sGRB jet, is a key but nontrivial aspect of our \textsc{pluto} simulations, where different physically motivated choices are possible. 
To address this issue, in the following Appendix\,\ref{IDEALvsTAUB} we compare simulation results obtained by adopting a Taub EOS and a much simpler $\mathrm{\Gamma_{ad}\!=\!4/3}$ ideal gas law. 
Then, in Appendix\,\ref{Mismatch}, we investigate instead the impact, for the Taub EOS, of importing initial data for specific internal energy instead of pressure from the outcome of the reference BNS merger simulation (see also Section\,\ref{import}), thereby accounting for the mismatch between the EOS adopted in \textsc{pluto} and the one employed during the reference simulation itself.  

\subsection{Taub vs. \texorpdfstring{$\boldsymbol{\mathrm{\Gamma_{ad}\!=\!4/3}}$}{gamma_43}}\label{IDEALvsTAUB}

In our fiducial model (see Section\,\ref{fiducial}), a Taub EOS is adopted to simultaneously describe the post-merger surrounding environment and the injected sGRB jet. This choice is justified by the fact that, within the density range of data import, the EOS employed in the reference BNS merger simulation (from which the environment was initially imported) is rather close to an ideal gas with $\mathrm{\Gamma_{ad}\!=\!5/3}$, which can be reproduced well by the Taub EOS. 
At the same time, such an EOS reproduces the ideal gas with $\mathrm{\Gamma_{ad}\!=\!4/3}$ required within the ultra-relativistic jet.  

Here we discuss the results of two additional \textsc{pluto} simulations that we performed, adopting an ideal gas EOS with $\mathrm{\Gamma_{ad}\!=\!4/3}$ from the time of data import (`$\mathrm{ideal-4/3}$' case), or from the time of jet injection (`mixed' case), respectively, to study the impact of the EOS choice on the overall system evolution.

\begin{figure}
    \includegraphics[width=0.8\columnwidth,keepaspectratio]{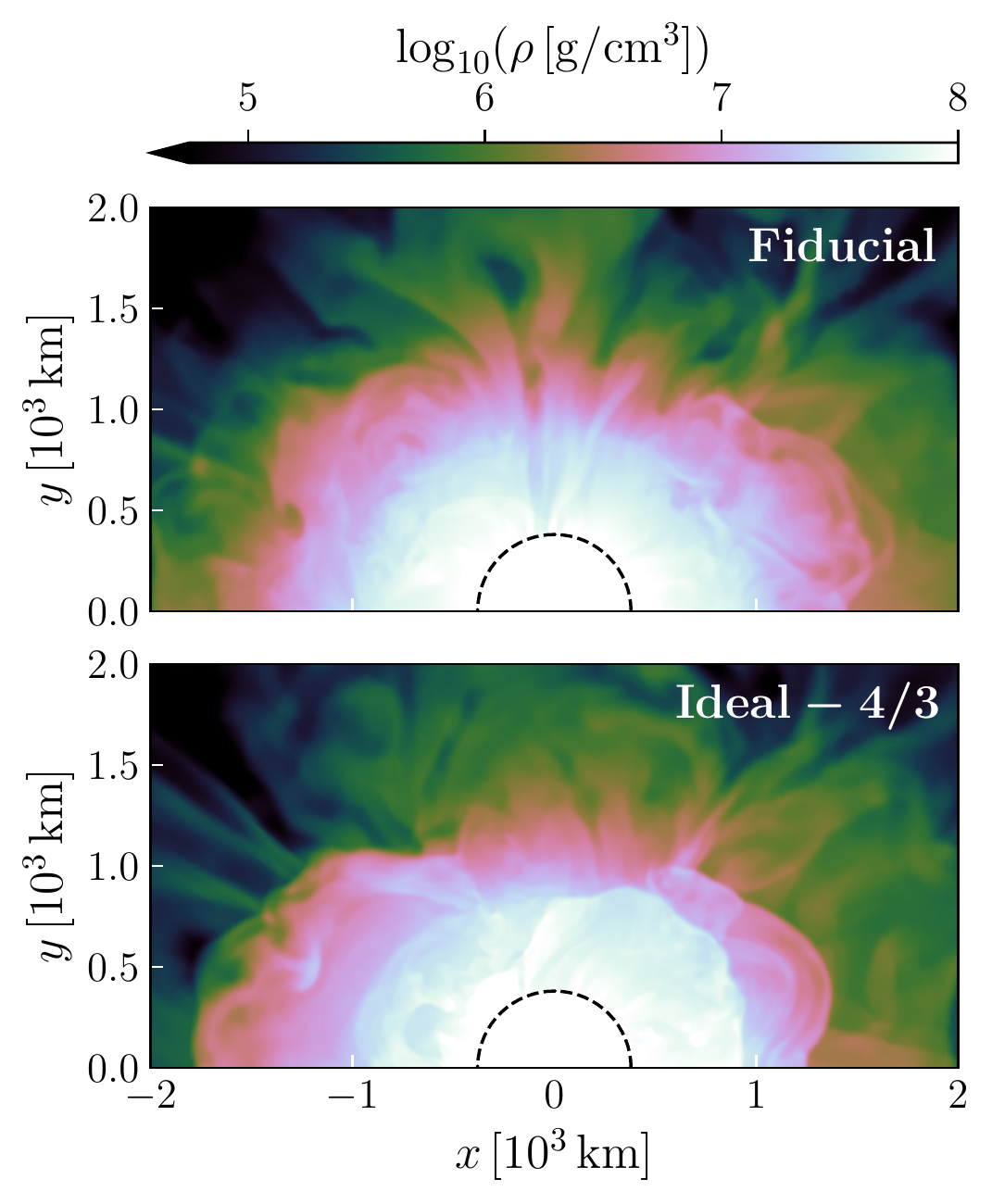}
    \caption{Meridional view of rest-mass density in the $xy$ plane ($\mathrm{\mathnormal{y}\!\ge\!0}$) at the time of jet injection (namely, $\mathrm{\simeq385\,ms}$ after merger). Top panel refers to our fiducial case, bottom panel to the case where an ideal gas EOS with $\mathrm{\Gamma_{ad}\!=\!4/3}$ is adopted.
    Dashed-black line indicates the inner radial boundary or excision radius. }
    \label{figC1}
\end{figure}

In Figure\,\ref{figC1}, we show the rest-mass density distribution in the $xy$ plane ($\mathrm{\mathnormal{y}\!\ge\!0}$) at the time of jet injection (i.e., $\mathrm{\simeq385\,ms}$ after merger), in the fiducial case (upper panel) and in the $\mathrm{ideal-4/3}$ case (lower panel). 
We recall that both density and pressure are directly set from the original BNS merger simulation at the initial data import ($t_0\simeq 155$\,ms after merger), and, up to 1800\,km radius, in two following substitution steps (see Section~\ref{substitution}). A different EOS adopted in \textsc{pluto} corresponds to a different internal energy density, which then leads to deviations in the overall evolution. 
From the comparison given in Figure~\ref{figC1}, relevant differences are clearly visible.
In particular, in the fiducial case we point out the presence of a lower-density funnel along the orbital $y$-axis of the system, excavated by the collimated outflow identified in Figure\,\ref{fig1}. Conversely, in the $\mathrm{ideal-4/3}$ case, we do not see such a distinctive feature of the system, wiped out by the excessive thermal energy resulting from such an EOS.

\begin{figure*}
	\includegraphics[width=1.4\columnwidth,keepaspectratio]{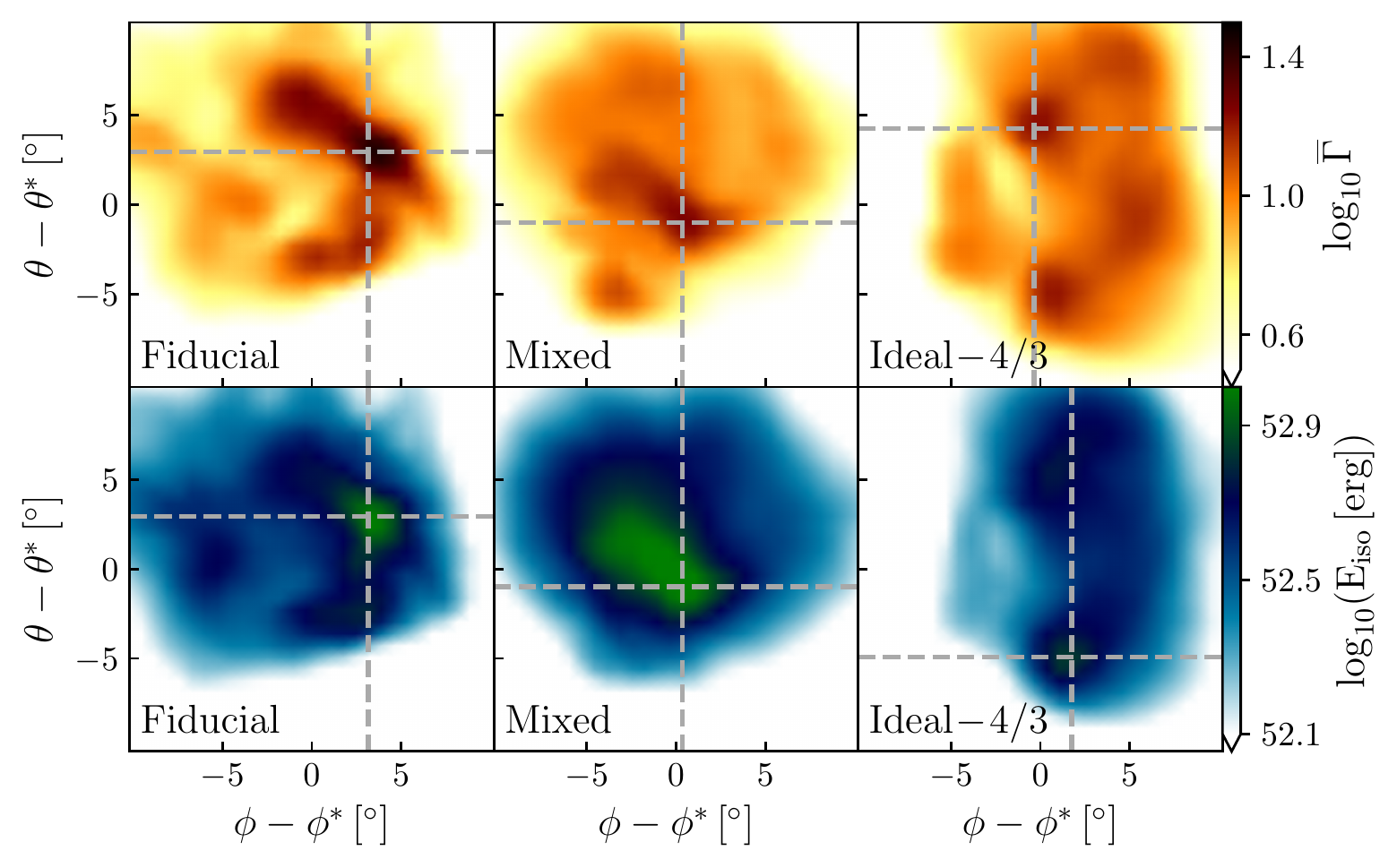}
    \caption{2D angular profiles of the northern head of the jet ($r\!\in\!2-6\times10^5$\,km) at $\mathrm{\simeq\!2385\,ms}$ after merger, for the fiducial, mixed, and `$\mathrm{ideal-4/3}$' cases, respectively (see text). In the upper panels, we show the distribution of radial-averaged Lorentz factor (weighted over the total energy density), and in the lower panels the corresponding isotropic equivalent energy distribution. }
    \label{figC2}
\end{figure*}

In Figure\,\ref{figC2}, we illustrate the angular properties of the jet head (from $\mathrm{\sim\!2\!\times\!10^5\,km}$ to $\mathrm{\sim\!6\!\times\!10^5\,km}$, north pole only) at $\mathrm{\simeq\!2385\,ms}$ after merger. In particular, we show results of three different simulations, i.e., fiducial (left panels), mixed (center panels) and $\mathrm{ideal-4/3}$ (right panels), displaying the distributions of radial-averaged Lorentz factor (top), and isotropic equivalent energy (bottom). 
First, we notice that differences in $E_\mathrm{iso}$ between the left and center panels are much smaller than those with the right panels.
This suggests that the EOS choice primarily impacts the evolution of the environment into which the jet is injected, with indirect but large effects on the subsequent jet propagation process. As pointed out earlier, assuming from the beginning a gas pressure dominated by radiation (i.e., $\mathrm{\Gamma_{ad}\!=\!4/3}$) leads in particular to the absence of a lower-density funnel along the injection axis (Figure\,\ref{figC1}), causing in turn significant jet choking at later stages (Figure\,\ref{figC2}, right panels).
On the other hand, if the different EOS is only introduced at the time of jet injection (with same starting environment), then the effects on the jet evolution are strongly reduced. 

Comparing the fiducial and mixed cases (left and central panels of Figure\,\ref{figC2}), we observe for the latter a larger degree of Gaussianity and axisymmetry of both $\overline{\Gamma}$ and $\mathrm{E_{iso}}$ distributions, which also result better centered on the $y$-axis ($\mathrm{\mathnormal{\theta}\!=\!\mathnormal{\theta}^*,\,\mathnormal{\phi}\!=\!\mathnormal{\phi}^*}$).
Maxima values of $\overline{\Gamma}$ and $\mathrm{E_{iso}}$ remain however consistent with those of the fiducial case.\footnote{The absence of major differences in jet propagation when using a Taub EOS or a $\mathrm{\Gamma_{ad}\!=\!4/3}$ EOS (with same starting environment) was pointed out also in \cite{Harrison2018}, where the authors simulated with \textsc{pluto} the case of a 2D unmagnetized jet propagating in a hand-made medium. Here we extend similar conclusions to the case of 3D magnetized jets with realistic injection environments in BNS merger context.}
As a note of caution, we stress again that the EOS in the original BNS merger simulation is closer to  $\mathrm{\Gamma_{ad}\!=\!5/3}$ and therefore the higher axisymmetry and Gaussianity found with 4/3 may be partially due to the introduction of more significant EOS deviations (for what concerns the environment). 

As mentioned in Section\,\ref{import}, a Taub EOS has the disadvantage of not allowing for a consistent reproduction of radiation-mediated shocks formed during the propagation of sGRB jets, for which an ideal gas with $\mathrm{\Gamma_{ad}\!=\!4/3}$ seems more appropriate (e.g., \citealt{Gottlieb2022a}). 
A way to obtain more consistent results, would be not only to impose the latter in \textsc{pluto} simulations, but also adapt the EOS employed in future BNS merger simulations to be as close as possible to an ideal gas with $\mathrm{\Gamma_{ad}\!=\!4/3}$ below a reference density (which could be, e.g., of order $10^8$\,g/cm$^3$).

\subsection{EOS mismatch with reference BNS merger simulation}\label{Mismatch}

During our import and substitution steps (Sections\,\ref{import} and \ref{substitution}, respectively), data for the rest-mass density and pressure were directly taken from the outcome of the reference BNS merger simulation, while the corresponding specific internal energy was computed via the Taub EOS. Such an EOS, however, does not exactly correspond to one employed in the reference merger simulation (within the density range of data import; see above), leading to a mismatch of our physical system (in terms of specific internal energy) with the original post-merger one.

As already pointed out in P21, one way to quantify the impact of the above mismatch on the overall system evolution reproduced by \textsc{pluto} is to compare the results of our fiducial model obtained by the above steps (also see Section\,\ref{fiducial}) with those obtained by using the same numerical and physical setup, but importing the specific internal energy from the outcome of the reference BNS merger simulation and then calculating the corresponding pressure via the Taub EOS (model referred to as `$\epsilon\_\textsc{bns}$' in what follows).

\begin{figure}
    \includegraphics[width=\columnwidth,keepaspectratio]{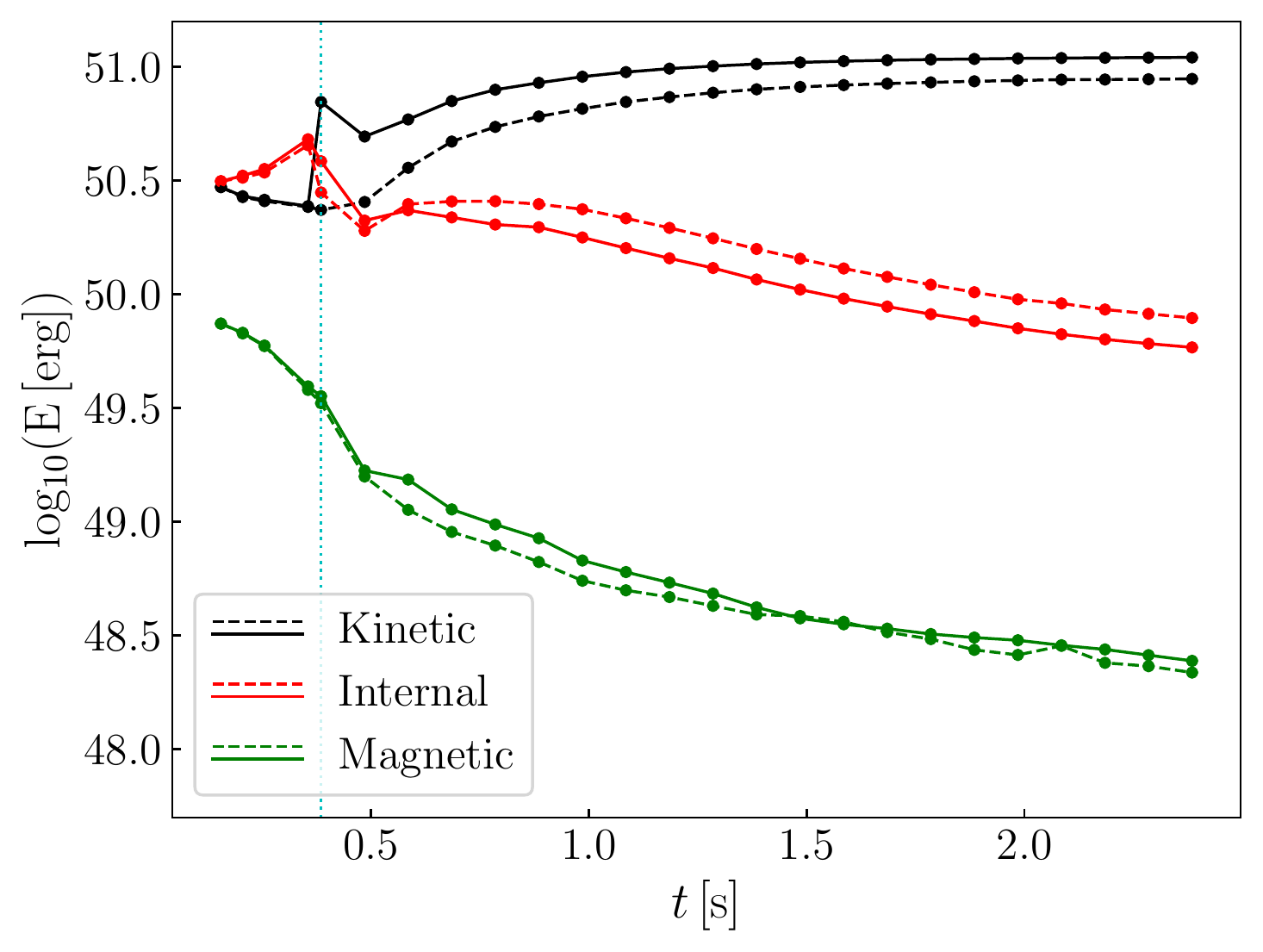}
    \caption{Evolution of kinetic, internal, and magnetic energies in the fiducial (solid lines) and $\epsilon\_\textsc{bns}$ (dashed lines) models (see text). The energy values are computed over the entire computational domain and at different simulation times, from 155\,ms after merger (data import time) to the end of the simulation (2385\,ms after merger). The vertical cyan dotted line marks the end of the collapse phase (385\,ms after merger).}
    \label{figC3}
\end{figure}

In Figure\,\ref{figC3} we compare the two cases (direct import of pressure vs.~direct import of specific internal energy) in terms of the total kinetic (black), internal (red), and magnetic (green) energies, computed over the entire computational domain from the time of data import (155\,ms after merger) to 2\,s after jet launch (namely, 2385\,ms after merger).
In the Figure, solid and dashed lines correspond to the fiducial and the $\epsilon\_\textsc{bns}$ model, respectively.

Up to the beginning of the collapse phase, i.e.~355\,ms after merger (fourth point from the left in each energy profile), the two cases only differ by at most $\simeq\!6\%$. During the collapse phase, however, the relative differences in kinetic and internal energies increase considerably, reaching a maximum of $\simeq\!66\%$ and $\simeq\!27\%$, respectively, at the end of the same phase (vertical cyan dotted line in the Figure). The relative difference in magnetic energy instead remains stable around $7\%$. 
The sudden increase of the discrepancy in kinetic and internal energies turns out to be connected with the zero-gradient radial boundary conditions imposed during the collapse phase at $r\!=\!r_\mathrm{exc}$\,($=\!380$\,km), which, due to the differences in pressure and specific internal energy between the two simulations, lead to a significantly different physical flux on the corresponding spherical surface. The aforementioned increase in fact takes place only within a radius of $\simeq\!10^3$\,km ($\gtrsim\!r_\mathrm{exc}$), and starting at 375\,ms after merger, when the system already extends up to $\simeq\!1.5\times10^4$\,km ($\gg\!r_\mathrm{exc}$). 

The correspondence in magnetic energy between the two models until the end of the collapse phase indicates that the differences in kinetic and internal energies are not connected with the presence of magnetic fields. This agrees with the fact that only the thermodynamic properties of the system are affected by the change of EOS with respect to the reference BNS merger simulation. However, we note that compared with P21, wherein magnetic fields where not taken into account at all, the above differences in kinetic and internal energies are now much more pronounced (in P21 they were only a few percent). 
This is probably due to the longer duration of the collapse phase itself (20\,ms longer than in P21), which results in the amplification, during the dynamical evolution, of effects associated with the zero-gradient boundary conditions set at $r\!=\!r_\mathrm{exc}$.    

After the injection of the sGRB jet, carried out using the numerical prescription detailed in Section\,\ref{jet}, we see a significant reduction in the relative difference in kinetic energy, which keeps decreasing and becomes $\simeq\!20\%$ at the end of the simulation. This is due to the fact that the contribution from the jet, common to the two cases, dominates more and more over the environment.
On the other hand, the early interaction of the jet with the surrounding environment, which contains most of the internal energy of the system (see Figure\,\ref{fig9}), leads to a discrepancy of $\simeq\!30\%$ in the latter, which persists in the following evolution. 
A smaller discrepancy of $\simeq\!10\%$ emerges also in the magnetic energy, although this seems to decrease again as the jet evolves. 

The results shown in Figure\,\ref{figC3}, in conclusion, indicate that the mismatch between the EOS chosen in \textsc{pluto} and the one employed in the reference BNS merger simulation has a non-negligible impact on the system dynamical evolution. Notably, this impact is more severe than that found in P21, given the longer duration of the collapse phase assumed in the present work. 

As pointed out at the end of Appendix\,\ref{IDEALvsTAUB}, employing in future reference BNS merger simulations the same EOS used in the following \textsc{pluto} simulations, within the density range of data import (i.e.~$\lesssim\!10^8\,\mathrm{g/cm^3}$), will allow us to achieve fully consistent results and eliminate any form of mismatch.


\bsp	
\label{lastpage}
\end{document}